\documentclass[structabstract]{aa}  

\usepackage{graphicx}
\usepackage{txfonts}

\usepackage[table]{xcolor}

\RequirePackage{array,hyperref} 
\RequirePackage{subfig,color,pifont}
\RequirePackage{blkarray,hhline,multirow}
\RequirePackage{bold-extra}
\RequirePackage{bbm}
\RequirePackage{etoolbox}
\RequirePackage{rotating}

\usepackage{empheq}

\definecolor{dkmauve}{rgb}{0.70,0.01,0.50}
\definecolor{dkblue}{rgb}{0.05,0.05,0.70}

\hypersetup{ bookmarks=true, unicode=true, pdftoolbar=true, pdfmenubar=true, 
	pdffitwindow=true, pdfstartview={FitH}, pdfauthor={SP}, 
	colorlinks=true, linkcolor=dkblue,citecolor=dkblue, urlcolor=dkmauve, breaklinks }

\begin{document}

\title{High-redshift quasar selection from the CFHQSIR survey}

\author{S. Pipien \inst{1}
\and J.-G. Cuby \inst{1}
\and S. Basa \inst{1}
\and C. J. Willott \inst{2}
\and J.-C. Cuillandre \inst{3}
\and S. Arnouts \inst{1}
\and P. Hudelot \inst{4}
}
\institute{Aix Marseille Univ, CNRS, CNES, LAM, Marseille, France\\
              \email{sarah.pipien@lam.fr}
\and
NRC Herzberg, 5071 West Saanich Rd, Victoria, BC V9E 2E7, Canada
\and
CEA/IRFU/SAp, Laboratoire AIM Paris-Saclay, CNRS/INSU, Université Paris Diderot, Observatoire de Paris, PSL Research University, F-91191 Gif-sur-Yvette Cedex, France
\and
Institut d’Astrophysique de Paris, 98bis Boulevard Arago, F-75014 PARIS, France\\
}

\abstract{
Being observed only one billion years after the Big Bang, $z \sim 7$ quasars are a unique opportunity for exploring the early Universe. However, only two $z \sim 7$ quasars have been discovered in near-infrared surveys: the quasars ULAS J1120+0641 and ULAS J1342+0928 at $z = 7.09$ and $z = 7.54$, respectively. The rarity of these distant objects, combined with the difficulty of distinguishing them from the much more numerous population of Galactic low-mass stars, requires using efficient selection procedures. The Canada-France High-z Quasar Survey in the Near Infrared (CFHQSIR) has been carried out to search for $z \sim 7$ quasars using near-infrared and optical imaging from the Canada-France Hawaii Telescope (CFHT). Our data consist of $\rm{\sim 130\,deg^{2}}$ of Wide-field Infrared Camera (WIRCam) Y-band images up to a $5\,\sigma$ limit of $\rm{Y_{AB} \sim 22.4}$ distributed over the Canada-France-Hawaii Telescope Legacy Survey (CFHTLS) Wide fields. After follow-up observations in J band, a first photometric selection based on simple colour criteria led us to identify 36 sources with measured high-redshift quasar colours. However, we expect to detect only $\sim$ 2 quasars in the redshift range $6.8 < z < 7.5$ down to a rest-frame absolute magnitude of $\rm{M_{1450}} = -24.6$. With the motivation of ranking our high-redshift quasar candidates in the best possible way, we developed an advanced classification method based on Bayesian formalism in which we model the high-redshift quasars and low-mass star populations. The model includes the colour diversity of the two populations and the variation in space density of the low-mass stars with Galactic latitude, and it is combined with our observational data. 
For each candidate, we compute the probability of being a high-redshift quasar rather than a low-mass star. This results in a refined list of the most promising candidates. Our Bayesian selection procedure has proven to be a powerful technique for identifying the best candidates of any photometrically selected sample of objects, and it is easily extendable to other surveys.}


\maketitle
\titlerunning{CFHQSIR}

\section{Introduction}
Quasars reside at the centres of active galactic nuclei (AGNs) and are believed to be powered by mass accretion onto a supermassive black hole (SMBH). Thanks to their strong intrinsic luminosity, quasars are bright enough \citep[$\rm{L\sim10^{14} L_{\odot}}$ for the most luminous quasar ever discovered at $z = 6.30$, ][]{Wu2015} to be detected at high redshifts, where they can be used to explore the early Universe. Along with other cosmological probes, high-redshift quasars (hereafter, "high-redshift" refers to redshifts $z \gtrsim 5.6$) have proved to be powerful tools for studying not only the epoch of cosmic reionisation \citep[e.g.][]{Fan2006, Jiang2008, Becker2015}, but also for investigating the formation and evolution of primordial SMBHs \citep[e.g.][]{Willott20102}. 

The relations between luminosity, black hole mass, and broad emission line velocity have been demonstrated at low redshifts by reverberation mapping studies \citep[e.g.][]{Kaspi2000, Vestergaard2002, Bentz2009}. Spectroscopy of high-redshift quasars (high-z quasars) can therefore be exploited in order to measure SMBHs masses, estimate their mass function, and place solid constraints on SMBH formation models, under the assumption that these local relations are still valid at high redshifts. Using this technique, it has been possible to confirm black holes with masses exceeding $\rm{M_{BH}} \gtrsim 10^{9} \rm{M_{\odot}}$ at redshifts $z \gtrsim 6$ \citep{Mortlock2011, Wu2015, Banados2018}. How such massive objects could have formed so quickly in less than 1 Gyr after the Big Bang is still a fundamental question arising from these discoveries \citep[see reviews by][]{Volonteri2010, Haiman2013, Smith2017}. Several scenarios for the formation of SMBHs seeds, including, for instance, remnants of Population III stars \citep[e.g.][]{Madau2001, Volonteri2003} or a direct collapse of gas in atomic cooling halos \citep[DCBH, e.g.][]{Visbal2014, Smidt2017} have been proposed, but they are widely debated, partly because known bright high-z quasars are likely to be the tip of an iceberg that is mainly composed of fainter quasars.

The cosmic reionisation that ended the so-called dark ages during which the first sources of the Universe ionised the hydrogen in the intergalactic medium (IGM) is a major event in the Universe's history, and many questions related to the onset, duration, topology, and the sources responsible for this process remain unsolved. High-redshift quasars can be used as background sources whose UV radiation is absorbed at the resonant Lyman lines by intervening neutral hydrogen along the line of sight. Their spectra are therefore reliable tools for absorption studies since they show spectral signatures of the IGM state. The hydrogen neutral fraction $x_{HI}$ can indeed be determined by making a range of measurement on high-z quasar spectra, such as the Gunn-Peterson test \citep[e.g.][]{Gunn1965, Fan2006, Becker2015b}, near-zone measurement \citep[e.g.][]{Venemans2015, Mazzucchelli2017}, dark gaps and dark pixels statistics \citep[e.g.][]{McGreer2015}, and Lyman-$\alpha$ damping wing reconstruction \citep[e.g.][]{Greig2017}. Measurements of the quasar luminosity function (QLF) at $z\sim6$ have shown that the quasar ionising flux was not sufficient to keep the Universe ionised, with a photon rate density lower by between 20 and 100 times than required \citep{Willott2010}. Even though the faint end of the QLF remains poorly constrained, it is now generally agreed that AGNs do not contribute significantly to the required ionising photon budget at $z \sim 6$ \citep{Onoue2017}. \\

If there are compelling reasons to search for high-z quasars, the quest is no less challenging because of their rarity: \citet{Willott2010} predicted a number of quasars of the order of $\rm{\sim 0.1\,deg^{-2}}$ brighter than $\rm{H_{AB} \simeq 24}$ in the redshift range $6.5 < z < 7.5$. In the past few years, substantial progress has been made in this field, where more than 100 high-z quasars have been identified \citep{Banados2016}. Wide area surveys greatly contributed to this success, with surveys first carried out in optical bands, such as the Sloan Digital Sky Survey \citep[SDSS ]{York2000} and the Canada-France High-z Quasar Survey \citep[CFHQS;][]{Willott2005, Willott2007, Willott2009, Willott2010}, providing more than 50 quasars up to redshifts $z \simeq 6.4$. At higher redshifts, the Lyman-$\alpha$ emission line is shifted into the near-infrared (NIR) and becomes undetectable at observed wavelengths $\rm{\lambda_{obs}} \lesssim 0.9\,\rm{\mu} m$ because of IGM absorption; this makes the use of near-IR bands necessary. Ongoing near-IR surveys such as the United Kingdom Infrared Telescope (UKIRT) Infrared Deep Sky Survey \citep[UKIDSS; ][]{Lawrence2007}, the Visible and Infrared Survey Telescope for Astronomy (VISTA) Kilo-degree Infrared Galaxy\citep[VIKING; ][]{Emerson2004}, the Panoramic Survey Telescope \& Rapid Response System \citep[Pan-STARRS; ][]{Kaiser2010}, the Dark Energy Survey \citep[DES; ][]{DES2016}, the VISTA Hemisphere Survey \citep[VHS; ][]{McMahon2013}, and the Subaru HSC-SSP Survey with the Subaru High-z Exploration of Low-Luminosity Quasars (SHELLQs) project \citep{Miyazaki2012, Matsuoka2016} have started to increase the number of known  $z \gtrsim 6.5$ quasars by employing filters centred on $\rm{1\,\mu m}$. The use of such data allowed \citet{Mortlock2011} to discover the previous redshift record holder, ULAS J1120+0641 at $z=7.09$, before it was superceded by ULAS J1342+0928 at $z=7.54,$ for which \citet{Banados2018} mined data from the UKIDSS Large Area Survey \citep{Lawrence2007}, the Wide-field Infrared Survey Explorer \citep[ALLWISE, ][]{Wright2010}, and the DECam Legacy Survey (DECaLS\footnote{\url{http://legacysurvey.org/decamls/}}). However, despite all these efforts, these two quasars are the only ones  found at redshift $z > 7$ to date. The Canada-France High-z Quasar Survey in the Near Infrared (CFHQSIR) has been designed to find more $z \sim 7$ quasars that can be used to constrain the reionisation epoch as well as the initial growth of SMBHs. \\
The inevitable contamination of any photometric sample by Galactic low-mass stars (main-sequence stars and brown dwarfs) increases the difficulty of finding such rare objects. Colour-colour cuts are commonly used for high-z quasars searches to separate the two populations of sources \citep[e.g.][]{Willott2005, Venemans2013}. However, the photometric noise implies that the resulting candidate list is likely to be still dominated by low-mass stars, which are much more numerous than high-z quasars. Here we adopt a powerful classification technique that is based on Bayesian inference. It was first developed by \citet{Mortlock2012} but was also applied by \citet{Matsuoka2016}, who discovered 39 quasars in the redshift range $5.7 \leq z \leq 6.9$ \citep{Matsuoka2016, Matsuoka2018}. The technique allows assigning to each candidate a probability of being a high-z quasar rather than a low-mass star by combining all the information (observations and prior knowledge) available for each source in an optimal way. Although this approach has not been frequently used, it has proved to be a powerful method since \citet{Mortlock2011} successfully found the second most distant quasar at redshift $z = 7.09$. \\
This paper describes an improved method for searching for $z\sim 7$ quasars, and we report our initial results. In the next section we present the CFHQSIR data and our colour selection criteria. In Sect. \ref{sec3} we describe our photometric candidate selection and detail our Bayesian method of classifying candidates. Our spectroscopic observations and initial results are presented in Sect. \ref{secresults}. Section \ref{sec5} summarises these results and presents our conclusions. All magnitudes in the optical and near-IR bands in this paper are in the AB system. Cosmological parameters of $\rm{H_{0}} = 68$ \mbox{km s$^{-1}$ Mpc$^{-1}$}, $\rm{\Omega _{M} = 0.31}$ and $\rm{\Omega _{\Lambda} = 0.69}$ are assumed throughout \citep{Planck2016}.

\section{Searching for high-z quasars in the CFHQSIR survey}
\label{sec2}

At high redshifts, neutral hydrogen in the IGM induces a complete absorption of the quasar flux blueward of the Ly-$\alpha$ line at $\rm{\lambda_{Ly\alpha}} = 1216\,\AA$. A $z \sim 7$ quasar has a clear photometric signature: it is relatively bright in the NIR above $\rm{1\,\mu m}$ and is totally extinguished in all optical bands. We therefore searched for $z \sim 7$ quasars through Y-band imaging of the CFHTLS Wide fields for which deep optical data exist. We briefly describe our CFHQSIR Y-band data in Sect. \ref{subcfhqsir}. We discuss our colour selection criteria in Sect. \ref{simulcolor} .

\subsection{CFHQSIR survey}
\label{subcfhqsir}

The CFHQSIR is a CFHT Large Program carried out at the 3.6m CFH telescope with the Wide field IR camera (WIRCam) from 2011 to 2013. It aims to extend the highly successful $5.8 < z < 6.5$ CFHQS survey \citep{Willott2005, Willott2007, Willott2009, Willott2010} to higher redshifts. The CFHQSIR covers $\sim 130$ deg$^2$ over the four CFHTLS Wide fields up to a $5\,\sigma$ limiting magnitude of $\mathrm{Y_{lim} \approx 22.4}$ for point sources and is characterised by an average image quality of 0.7\arcsec. The data were obtained in two epoch observations separated by at least 20 days in order to discard transient sources and slow-moving objects. We refer to \citet{Pipien} for further details about the CFHQSIR data. \\
Assuming the quasar luminosity function derived by \citet{Willott2010} at $z = 6$ and considering a pure density evolution with redshift, we estimated the expected number of high-z quasars in the CFHQSIR survey. About $2.3$ quasars are expected in the redshift range $6.78 < z < 7.48$ down to $\rm{M_{1450}} = -24.6$. A completeness correction in magnitude and redshift, as computed in \citet{Pipien} (Fig. 12) and in Sect. \ref{lowmass} (Fig. \ref{complredshift}), respectively, was applied to derive the predicted number. This number decreases from $2.3$ to $1.4$ when considering the quasar luminosity function and evolution from \citet{Jiang2016}, for which a steeper redshift evolution of the quasar density between $z \sim 5$ and $z \sim 6$ was demonstrated. Given this small number, it is therefore necessary to adopt a reliable selection procedure to efficiently identify and classify high-z quasar candidates. This is further described in Sects. \ref{simulcolor} and \ref{sec3}.

\subsection{Simulating the colours of quasars and low-mass stars}
\label{simulcolor}

Our selection of $z \sim 7$ quasar candidates is based on a combination of our CFHQSIR NIR (Y band) and optical data from the 7th and final release (T0007) of the CFHTLS (u-, g-, r-, i-, z- bands) produced by Terapix. At redshifts $z \gtrsim 6.0$, the strong break across the Ly-$\alpha$ line known as the Gunn-Peterson effect makes colour selection the most efficient way for selecting high-z quasars over large areas. At redshifts $z \gtrsim 6.5$, quasars can therefore be selected as z-band dropouts as their spectrum is totally extinguished in the u, g, r, and i bands. As demonstrated by \citet{Venemans2013} for the VIKING survey and by \citet{Mortlock2011} for UKIDSS, the most efficient way of distinguishing  high-z quasar and low-mass stars, their main source of contamination, is to use J-band photometry. In order to optimise our colour selection criteria, we performed simulations of  the expected optical and near-IR colours of quasars and low-mass stars as measured with the CFHT MegaCam (z filter) and WIRCam cameras (Y and J filters). The simulations are detailed in the following subsections. 

\subsubsection{High-redshift quasars}
\label{simulqso}
 
\begin{figure}\centering
        \includegraphics[width=0.45\textwidth]{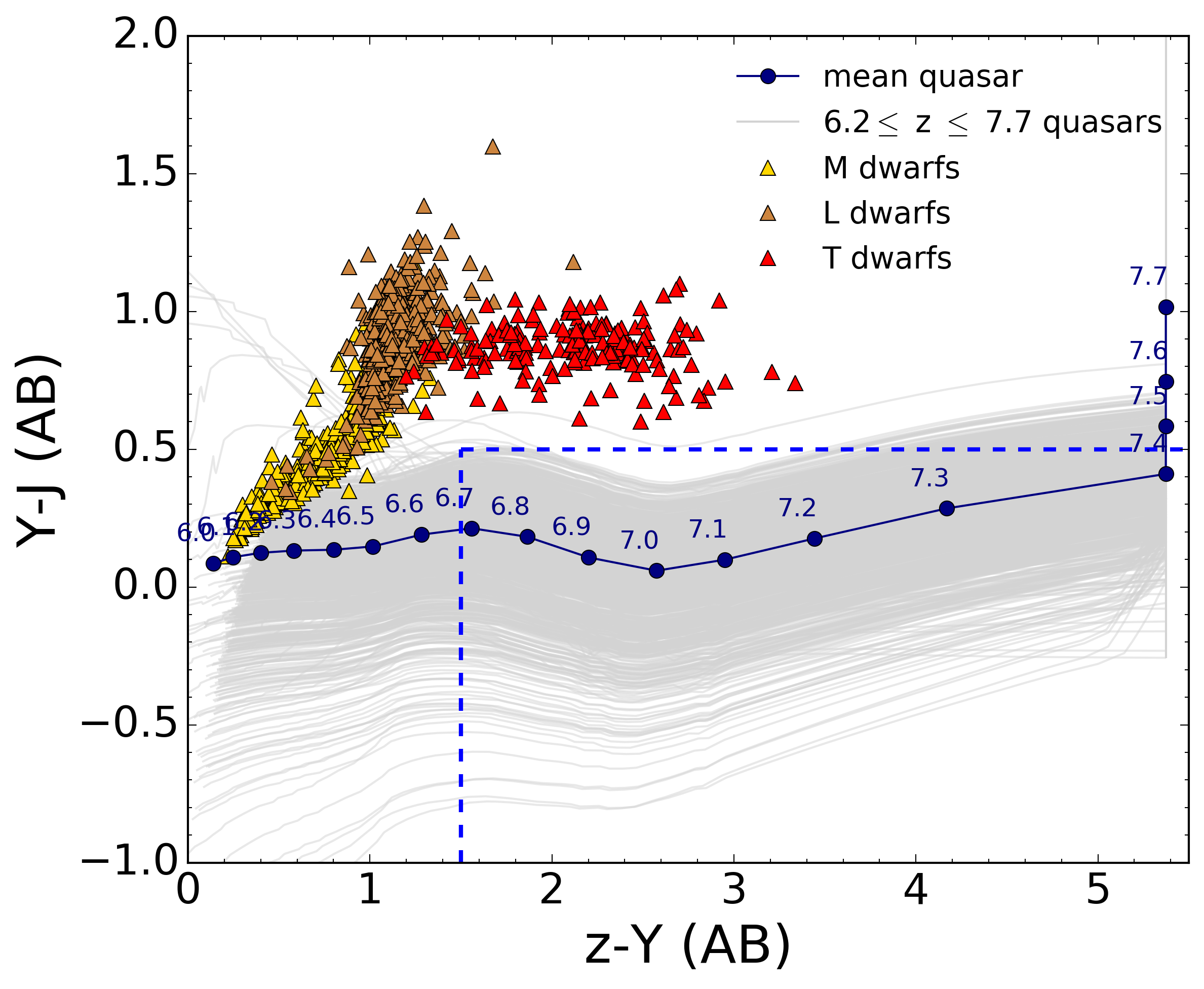}
        \caption{Diagram of $\rm{(z-Y)}$ vs. $\rm{(Y-J)}$  in the CFHT filters for simulated quasars and observed low-mass stars from the SpeX Prism library (triangles in yellow, brown, and red show M, L, and T dwarfs, respectively). The blue curve represents the colours of the mean quasar derived by \citet{Paris_2011} for which IGM absorption is included. The grey curves show the tracks in colour-colour space of the 1\,000 quasars redshifted from $z = 6.2$ to $z=7.7$. The blue dashed lines correspond to the colour criteria chosen to select quasars of redshift $z \sim 7$.}
        \label{colorsdiag}
\end{figure}

Based on a principal component analysis (PCA) of 78 $z\sim3$ quasar spectra of the SDSS-DR7, \cite{Paris_2011} derived the principal components characterising the quasar continuum and the PCA eigenvectors. We used the reported values of \citet{Paris_2011} (Table 2) of the distribution of their weights to simulate quasar mock spectra. We decided to consider only the first four components since $95\,\%$ of the variance affecting the 78 quasar spectra are represented by these components (I. Pâris, private communication). We assumed that the spectral index,  equivalent width, and intensity of the quasar emission lines at redshifts $z \sim 3$ are representative of the $z \sim 7$ quasar population. The intrinsic quasar luminosity plays an important role in validating this hypothesis since this quantity is anti-correlated with the line equivalent widths \citep[Baldwin effect, ][]{Baldwin1977}. As observed with the two most distant quasars ever discovered at redshift 7.1 and 7.5 \citep{Mortlock2011, Banados2018}, quasar spectra at low- and high-z match remarkably well: the intrinsic luminosities of high-z quasars coincide with those observed at lower redshifts and thus justify the use of $z\sim 3$ quasar spectra. We generated 1\,000 rest-frame quasar spectra, redshifted them in the redshift range $6.2 \leq z \leq 7.6$, and introduced IGM absorption assuming the transmission function of \citet{Meiksin2006}, where the transmitted flux at $ z \gtrsim 6.5$ was set to zero blueward of the Ly-$\alpha$ line due to the Gunn-Peterson trough. We then convolved the 1\,000 simulated quasar spectra with the z (MegaCam), Y, and J (WIRCam) CFHT filters including the atmospheric transmission. Fig. \ref{colorsdiag} shows the $\rm{(z-Y)}$ and $\rm{(Y-J)}$ colours obtained for all simulated quasar spectra. As expected, the quasar track shows a strong evolution of the $\rm{(z-Y)}$ colour with redshift, reflecting the movement of the Ly-$\alpha$ break through the z filter. At redshifts $z \lesssim 7.0$, the Ly-$\alpha$ emission line is in the z band, and it enters the Y band at $z \sim 7$. At higher redshifts, quasars become rapidly fainter in the z band and therefore redder in $\rm{(z-Y)}$. At redshifts $z \gtrsim 7.4$, the Ly-$\alpha$ break is redward of the z filter so that these quasars are entirely extinguished in this band. 

\subsubsection{Low-mass stars}
\label{lowmass}
We determined the expected colours of main-sequence stars and brown dwarfs (M, L, and T spectral types) based on a sample of observed spectra from the SpeX Prism library\footnote{The SpeX Prism Library and the SpeX Prism Library Analysis Toolkit are maintained by Adam Burgasser at \url{http://www.browndwarfs.org/spexprism}}. Only spectra covering the z (MegaCam), Y, and J (WIRCam) filters were used. More than 1\,300 spectra of sources with spectral types from M3 to T9 were exploited. The resulting colours computed in these filters are shown in Fig. \ref{colorsdiag}.  M and L-dwarfs represent a minor source of contamination compared to the much cooler T-dwarf population, which shows very similar $\rm{(z-Y)}$ colours to $z \sim 7$ quasars. However, the spectral distribution of T-dwarf peaks in the J band, leading to a significantly redder $\rm{(Y-J)}$ colour than quasars and thus allows a clear separation between the two populations of objects. \\
The box bounded by the blue dashed lines in Fig. \ref{colorsdiag} indicates our colour-colour high-z quasar selection criteria: 
\begin{empheq}{align}
\label{eqcolor}
\begin{cases}
\rm{z-Y} \geq 1.5 \\
\rm{Y-J} \leq 0.5
\end{cases}
\end{empheq}

\subsubsection{Colour-selection completeness}

\begin{figure}\centering
        \includegraphics[width=0.45\textwidth]{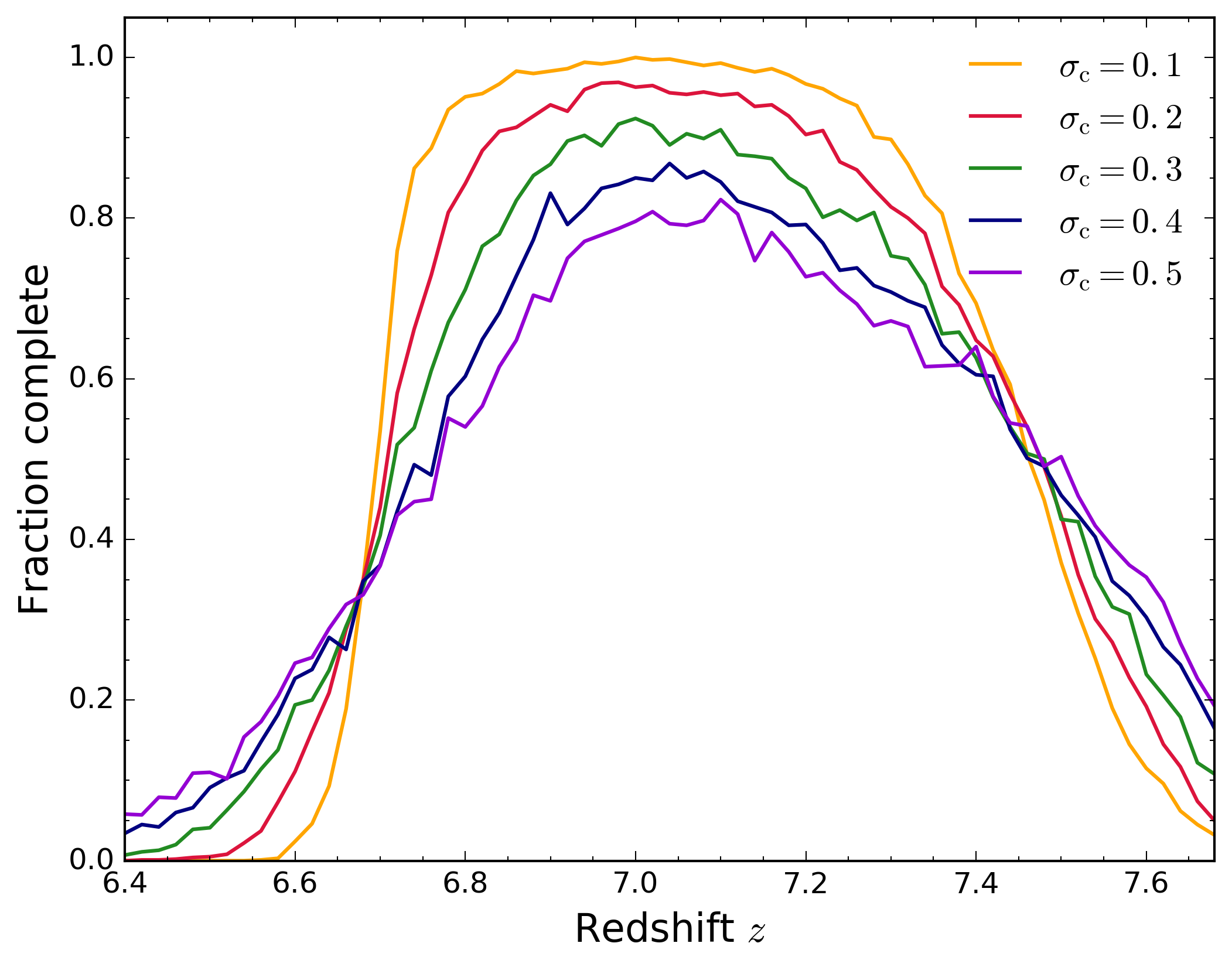}
        \caption{Fraction of high-z quasars satisfying the colour selection criteria of Eq. \ref{eqcolor} as a function of redshift. Different photometric uncertainties on $\rm{(z-Y)}$ and $\rm{(Y-J)}$ are taken into account, from $\sigma_{c} = 0.1$ to $\sigma_{c} = 0.5$. }
        \label{complredshift}
\end{figure}

To study the quasar selection efficiency as a function of redshift more precisely, we followed the approach developed by \citet{Willott2005} by calculating the fraction of artificial quasars that satisfies the criteria of Eq. \ref{eqcolor} as a function of redshift. In order to take into account photometric errors affecting the measurement of the $\rm{(z-Y)}$ and $\rm{(Y-J)}$ colours, each simulated quasar colour was described by a Gaussian probability density distribution with a standard deviation noted $\sigma_{c}$. We note, however, that as discussed in Sect. \ref{bayesian}, it would be more accurate to model each Gaussian in flux units rather than in magnitude units. Figure \ref{complredshift} shows the resulting fraction of high-z quasars falling into the colour-colour box delimited by Eq. \ref{eqcolor}. The different curves illustrate the change in completeness as a function of the error in colours. To take into account the correlated noise, each photometric error was multiplied by a corrective excess noise factor derived for each image \citep[see][for further details]{Pipien}. Figure \ref{complredshift} allows us to define the redshift range probed by our survey for a given completeness. Adopting $\sigma_{c} = 0.5$ as representative of the errors in colours for our sources, the CFHQSIR redshift range corresponding to a completeness greater than 50\% is $6.78 < z < 7.48$.

\section{Photometric candidate selection and Bayesian classification}
\label{sec3}
\subsection{Initial candidate selection}

\begin{figure*}[ht!]
        \centering
        \includegraphics[width=0.85\textwidth]{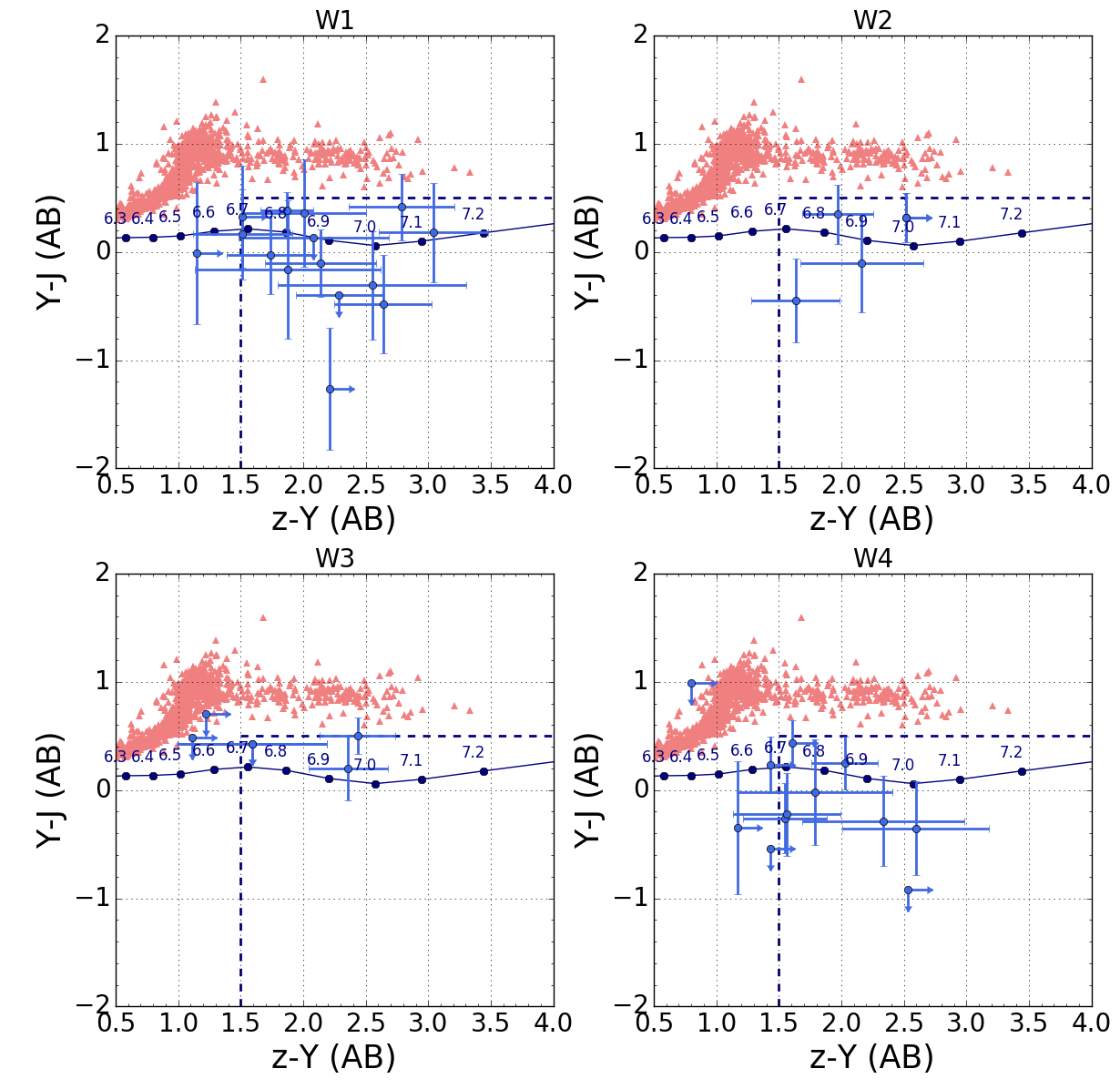}
        \caption{Colour-colour diagram showing the 36 high-z quasar candidates (blue points) identified in the four CFHTLS Wide fields. The arrow points correspond to a $5\,\sigma$ magnitude lower limit in the case of a non-detection in the z and/or J band.}
        \label{diag_cands}
\end{figure*}

Using the SExtractor software \citep{Bertin1996}, we produced catalogues of objects detected in the Y-band images. We first ran SExtractor in dual-image mode for source detection in Y-band images and photometry in all bands (u, g, r, i, z, and Y). We set the detection threshold to 1.5$\,\sigma$ above the background and required at least four connected pixels for a measurement. This double analysis of our near-IR and optical data reveals a total of $\sim$ 6 million sources detected in Y band. We then applied some automatic cuts including geometric and photometric criteria in order to keep the reddest point-like sources and remove the faintest objects in Y band. We set the following criteria: no detections in the u, g, r, and i filters, $\rm{(z-Y)}$ > 1.0, $Y \leq 23.0$ ($\rm{M_{1450}\lesssim}-$24.0), and a signal-to-noise ratio  $\rm{(S/N)_{Y}} \gtrsim 4$. All magnitudes were measured using the SExtractor \texttt{MAG\_AUTO} parameter. In total, $\sim$ 54\,000 objects satisfied these criteria and were selected. They were then inspected by eye in order to discard artefacts. We note that the main source of non-physical contamination comes from persistences. These effects occur in IR detectors when a saturated star observed in a previous exposure leaves a remnant image in the following exposures. These spurious sources appear in two different forms in our Y-band images. When the persistence is generated during an exposure with no connection with our observations, a single spot appears in our Y-band image. Since our observations were split into two epochs separated by at least 20 days, these spurious sources were easily spotted by inspecting our individual Y-band images before stacking \citep[for more details, see][]{Pipien}. The second form of persistence contaminating our data arises directly during our observations between two dithered exposures. This type of persistence is also easily identified since Y-band images show bright spots aligned according to the dither pattern. After eliminating these artefacts from our candidate list, we repeated the photometric measurements using SExtractor in single-image mode, which optimises the detection threshold and recentres the apertures. This process allowed us to remove sources with proper motion. We then again applied the same colour criteria as previously ($\rm{(z-Y)}$ > 1.0 and no detection in the u, g, r, and i filters) with these new measurements and selected 228 sources in this way. As discussed in Sect. \ref{sec2}, J-band photometry is necessary to distinguish high-z quasar candidates from low-mass stars. J-band follow-up observations were carried out at CFHT using the WIRCam camera, but also the SOFI (son of Isaac) and LIRIS (Long-slit Intermediate Resolution Infrared Spectrograph) cameras mounted on the 3.6m New Technology Telescope (ESO NTT) and the 4.2m William Herschel Telescope (WHT), respectively. The detectors used by SOFI and LIRIS are HgCdTe 1024 $\times$ 1024 Hawaii arrays with pixel scales of 0\arcsec.288. and 0\arcsec.251, respectively. LIRIS data were reduced using the Image Reduction and Analysis Facility (IRAF\footnote{IRAF is distributed by the National Optical Astronomy Observatory, which is operated by the Association of Universities for Research in Astronomy (AURA) under a cooperative agreement with the National Science Foundation.}), and SOFI images were reduced with the SOFI pipeline\footnote{\url{https://www.eso.org/sci/facilities/lasilla/instruments/sofi/tools/SofI_Pipeline.html}}. The photometric calibration for both datasets was performed following a method similar to that presented in \citet{Pipien}. We astrometrically calibrated the LIRIS images with the SCAMP\footnote{\url{https://www.astromatic.net/pubsvn/software/scamp/trunk/doc/scamp.pdf}} software and used the on-line service \href{http://nova.astrometry.net/}{Astrometry.net} for SOFI images.\\
We combined our follow-up observations with existing data when available, coming from the Canada France Brown Dwarfs Survey \citep[CFBDSIR, J band; ][]{Delorme2008}, the Visible and Infrared Survey Telescope for Astronomy \citep[VISTA-VHS, J-, H-, and $\rm{K_{s}}$-bands; ][]{McMahon2013}, and the Vimos  Public Extragalactic Redshift Survey \citep[VIPERS-Multi-Lambda Survey, $\rm{K_{s}}$ band; ][]{Moutard2016}. This allowed us to apply the colour cuts defined in Eq. \ref{eqcolor} and to confirm the reality of these objects. We identified a total of 36 high-z quasar candidates that are represented in the $\rm{(z-Y)}$ versus $\rm{(Y-J)}$ diagram of Fig. \ref{diag_cands}. Magnitudes were measured in 2.5\arcsec apertures, and aperture corrections were applied to give the total magnitudes. Non-detected objects in the z and/or J band (arrow points in Fig. \ref{diag_cands}) show colours outside the dashed selection region. The colour cuts defined in Eq. \ref{eqcolor} were not applied to these sources, as they only have a $5\,\sigma$ magnitude lower limit in these bands. Two sources in the W4 field showed blueer $\rm{(Y-J)}$ colours ($\rm{(Y-J)} < -0.5$) than the other candidates. Although they were not detected in the z or J filters, they are not spurious objects and remain candidates because they were both detected in the VIPERS $\rm{K_{s}}$ band. As shown in Fig. \ref{colorsdiag}, in the absence of photometric noise, high-z quasars and low-mass stars are expected to occupy a well-defined region in colour-colour space so that it is, in principle, possible to distinguish the two populations of objects by applying simple colour cuts. However, because high-z quasars are outnumbered by brown dwarfs and large photometric errors affect the faintest sources, this implies that most of our candidates are likely brown dwarfs scattered in the selection box defined in Eq. \ref{eqcolor}. In the next section, we present an alternative approach to the application of hard cuts based on Bayesian inference, which allowed us to classify our candidates in the best possible way according to their probability of being a high-z quasar rather than a brown dwarf. 

\subsection{Bayesian classification method}
\label{bayesian}

The Bayesian classification method allows introducing priors (e.g. the relative number of high-z quasars compared to the number of brown dwarfs), as opposed to the posterior distribution, which also includes observational data. The strength of this method is that a ranking of all candidates can be optimally established in a coherent statistical framework by combining all the information available for each object. In this section, we adapt and extend the Bayesian approach developed by \citet{Mortlock2012} to the CFHQSIR survey.\\
We assumed that the most important contamination comes from the low-mass stars and that artefacts and any other sources of contamination have been removed from our candidates list. We therefore considered only two population types: the high-z quasars, and the low-mass stars, for which we developed parametric models (see Sects. \ref{qsopop} and \ref{qsopop}).\\
Given a photometric dataset, we need to define the probability that a source detected in the CFHQSIR survey is a high-z quasar rather than a low-mass star. The answer is given by the conditional posterior probability, which can be estimated with the Bayes theorem by comparing parametric models for high-z quasars and low-mass stars. For a detected source with photometric measurements $\rm{\overrightarrow{d}}$, this probability is defined by
\begin{equation}
\rm{P_{q}=Pr(q|\overrightarrow{d})=\frac{W_{q}(\overrightarrow{d})}{W_{q}(\overrightarrow{d})+W_{s}(\overrightarrow{d})}}
\mathrm{,}
\end{equation}
where the weighted evidence terms are given by
\begin{equation}
\begin{multlined}
\rm{W_{q/s}(\overrightarrow{d})} \\
= \int \rm{\rho_{q/s}(\overrightarrow{p})}\:\rm{Pr(det\, |\, \overrightarrow{p}, q/s)}\:\rm{Pr(\overrightarrow{d}\, |\, \overrightarrow{p}, q/s)}\:\rm{d\overrightarrow{p}}.
\end{multlined}
\end{equation}
Here the subscripts $\rm{q}$ and $\rm{s}$ denote the high-z quasars and low-mass stars, respectively. The vector $\overrightarrow{p}$ corresponds to the parameters used to model the quasar and stars populations. The quantity $\rm{\rho_{q/s}(\overrightarrow{p})}$ represents the surface density of quasars or low-mass stars as a function of their intrinsic properties $\rm{\overrightarrow{p}}$. The functions $\rm{Pr(det\, |\, \overrightarrow{p}, q/s)}$ and $\rm{Pr(\overrightarrow{d}\, |\, \overrightarrow{p}, q/s)}$ are the probability that the source (quasar or star) is detected (det) and the probability that the source is observed with the photometric measurements $\rm{\overrightarrow{d}}$, each as a function of $\rm{\overrightarrow{p}}$.\\
As pointed out in \citet{Mortlock2012}, because the probability of a source to be a high-z quasar cannot be determined without at least some observational information (e.g. colour measurements),  the assumption is justified that the source has to be detected in the survey (in at least the Y band in our case). \\
This constraint does not, however, exclude the possibility that some candidates may be not detected in one or more photometric bands. This is especially important for high-z quasars, which can have negligible flux blueward of their Ly-$\alpha$ emission lines. Furthermore, a non-detection can lead to negative flux; this effect directly results from the photometric noise and is described in detail in \citet{Mortlock2012}. Given that such measurements cannot be converted into traditional logarithmic magnitudes \citep{Pogson1856}, we decided to follow the approach developed by \citet{Mortlock2012} by working in terms of flux units. In the following, for each population, stars or quasars, we therefore considered the observational dataset given by $\rm{\overrightarrow{d} = \left ( F_{z_{obs}}, F_{Y_{obs}}, F_{J_{obs}}  \right )}$, where $\rm{F_{z_{obs}}}$, $\rm{F_{Y_{obs}}}$, and $\rm{F_{J_{obs}}}$ are the observed flux of the candidates in the z, Y, and J bands, respectively. In the case of non-detections (in z or J), a flux is measured at the exact position where the object has been detected in Y band, using the SExtractor software in double-image mode.

\subsubsection{Quasar population}
\label{qsopop}

Our prior knowledge is given by the surface density of quasars $\rm{\rho_{q}(\overrightarrow{p})}$, which can be calculated with the quasar luminosity function of \citet{Willott2010}, noted $\Phi (\rm{M_{1450}},z)$. This function suggests the use of two parameters for the quasar population modelling: the rest-frame absolute magnitude $\rm{M_{1450}}$ , and the redshift $z$. However, we chose to work with the Y-band apparent magnitude instead of the absolute magnitude for practical reasons. The parameter set that describes the high-z quasar population is thus given by $\overrightarrow{p} = \left \{  \rm{Y_{mod}}, \textit{z}\right \}$, where $\rm{Y_{mod}}$ represents the modelled Y-band magnitude of the quasar. In the following, the subscript mod refers to fluxes and magnitudes estimated for the high-z quasar and low-mass star models. The surface density in \mbox{mag$^{-1}$ sr$^{-1}$ redshift$^{-1}$} can be written as a function of these two parameters:

\begin{equation}
\rm{\rho_{q}}(\rm{Y_{mod}},\textit{z}) = \frac{1}{4\pi} \times \frac{dV_{c}}{d\textit{z}} \times  \Phi \left [ \rm{Y_{mod}}-\mu - K_{corr}(\textit{z}),\textit{z} \right ] 
\mathrm{,}
\end{equation}

where $\mu$ is the distance modulus defined as a function of the luminosity distance $D_{L}$ by $\mu = 5log_{10}\left (  D_{L}/10\,\rm{pc}\right )$, $\rm{K_{corr}}(\textit{z})$ is the K-correction that converts the absolute magnitude at rest-frame 1450 $\AA$ into the observed Y-band magnitude, and $1/4\pi \times \rm{dV_{c}/d\textit{z}}$ is the comoving volume element per steradian and per redshift interval d$\textit{z}$.\\

For a given high-z quasar spectrum $\rm{q_{i}}$, the weighted evidence in \mbox{mag$^{-3}$ deg$^{-2}$} is

\begin{equation}
\label{Wqi}
\begin{multlined}
\rm{W_{q_{i}}}(\rm{z_{obs}}, \rm{Y_{obs}}, \rm{J_{obs}},\rm{det}) \\
= \int_{-\infty}^{+\infty} \int_{0}^{+\infty}\,\rho_{\rm{q_{i}}}(\rm{Y_{mod}},\textit{z})\,\rm{Pr(det | Y_{mod}}, \textit{z}, \rm{q_{i}}) \\ \times \rm{Pr(z_{obs}, Y_{obs}, J_{obs} | Y_{mod}}, \textit{z}, \rm{q_{i}) \,dY_{mod}}\,d\textit{z}.
\end{multlined}
\end{equation}

Since the Y band is the detection band of our survey, we modelled the probability $\rm{Pr(det | Y_{mod}, \textit{z}, q_{i})}$ with the completeness rate derived in \citet{Pipien}, which corresponds to a function varying between 0 and 1 and is characterised by a magnitude limit at 80 \% completeness of the order of $\rm{Y_{lim}} \sim 22.5$. \\
In the case of faint sources observed photometrically in the NIR, the Poissonian photon noise can reasonably be neglected. As most of our candidates are relatively faint (Y $\sim 21.0 - 23.0$), their photometric errors are dominated by background noise uncertainties, which are typically described by a normal distribution in flux. Ignoring the inter-band correlations, the probability $\rm{Pr(z_{obs}, Y_{obs}, J_{obs} | Y_{mod}, \textit{z}, q_{i})}$ is then given by a product of three Gaussians, each associated with one filter (z, Y or J), which, after conversion into magnitude units, is defined as 

\begin{equation}
\begin{multlined}
\label{vraisemblance}
\rm{Pr(z_{obs},Y_{obs}, J_{obs} | Y_{mod}, z, q_{i})} \\= \prod_{\rm{b}=1}^{\rm{N_{b}}=3} \frac{1}{\sqrt{2\pi}\sigma_{b}} \times \exp \left ( -\frac{1}{2} \left [ \frac{\rm{F_{b_{obs}}}-\rm{F_{b_{mod}}}(\overrightarrow{p})}{\sigma_{b}} \right ]^{2}\right ) \times \left | \frac{\rm{dF_{b_{mod}}}}{\rm{dm_{b_{mod}}}} \right | ,
\end{multlined}
\end{equation}

where the index b refers to the three bands considered here, b $= \left ( \rm{z, Y, J} \right)$, and $\rm{N_{b}}$ is the number of bands used. $\sigma_{b}$ represents the photometric errors following a Gaussian distribution in flux units. The Jacobian used to convert the probability from flux units into magnitude units is

\begin{equation}
\left | \frac{\rm{dF_{b_{mod}}}}{\rm{dm_{b_{mod}}}} \right | = \frac{2ln(10)\rm{F_{b_{mod}}}}{5}
\mathrm{,}
\end{equation}

where the model flux $\rm{F_{b_{mod}}}$ in the b band is given as a function of the zero-point flux noted $\rm{F_{0}}$,

\begin{equation}
\label{convertflux2}
\rm{F_{b_{mod}} = F_{0} e^{-2ln(10)m_{mod}/5}}.\\
\end{equation}

We refer to Appendix A2 of \citet{Mortlock2012} for a complete description of the magnitude - flux conversion for probability densities.\\

In Eq. \ref{Wqi}, the surface density $\rm{\rho_{q_{i}}(Y_{mod},\textit{z})}$ and the probability $\rm{Pr(z_{obs},Y_{obs}, J_{obs} | Y_{mod}, \textit{z}, q_{i})}$ are estimated for one single quasar template noted $\rm{q_{i}}$. We modelled the diversity of intrinsic properties of high-z quasars by considering 80 quasar templates, generated according to the method described in Sect. \ref{simulqso}. After assigning an equivalent weight to each quasar template $\rm{q_{i}}$, we write the final weighted evidence as the average of the evidences $\rm{W_{q_{i}}}$ over the $\rm{N_{temp}} = 80$ templates: 

\begin{equation}
\rm{W_{q}(z_{obs},Y_{obs}, J_{obs},det)} = \frac{1}{N_{temp}} \sum_{i=1}^{N_{temp}=80} \rm{W_{qi}(z_{obs}, Y_{obs}, J_{obs},det)}
\mathrm{.}
\end{equation}

\subsubsection{Brown dwarf population}
\label{starpop}

\begin{figure}[ht]\centering
	\includegraphics[width=0.45\textwidth]{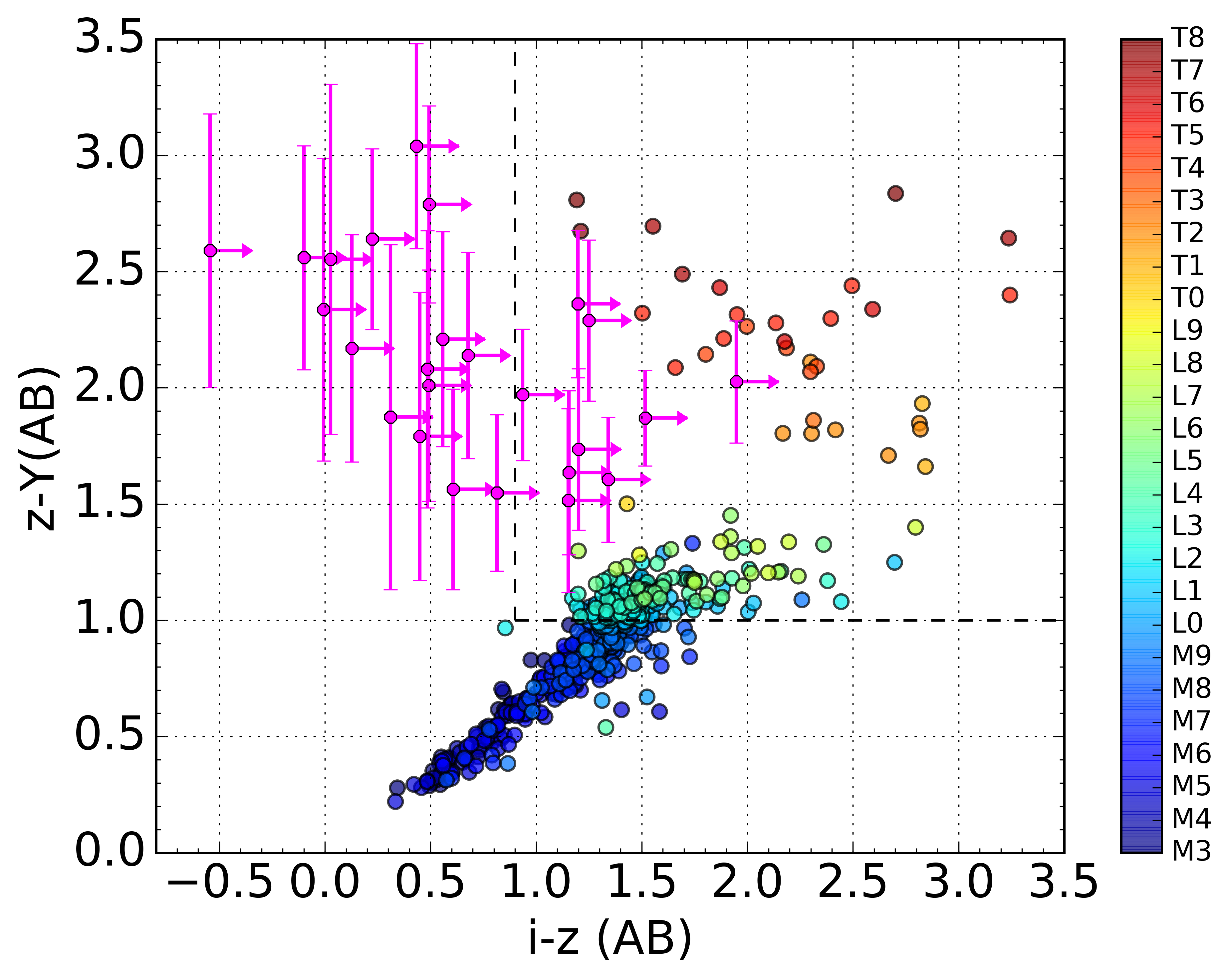}
	\caption{Observed colours of low-mass stars (spectra from the SpeX Prism library) of spectral types M3 (blue) to T8 (red). Colours of our high-z quasar candidates detected in the z band are also shown as magenta points. The box bounded by the dashed lines defines the spectral types that are likely to contaminate our candidate list.}
	\label{justifyBD}
\end{figure}

Given the colour criteria applied previously ($\rm{z-Y}$ $\geq$ 1.5, $\rm{Y-J}$ $\leq$ 0.5 and no detections in the u, g, r, and i filters), we determined the dominant source of contamination. To do this, we used a spectroscopic sample of M, L, and T stars from the SpeX Prism library covering the i, z, Y, and J bands. Most of the main-sequence stars are far bluer than high-z quasars, so that they do not represent a significant source of contamination. Conversely, T dwarfs are undetected in the i band, have $\rm{z-Y}$ colours mimicking $z \sim 7$ quasars, and noise can scatter their $\rm{Y-J}$ colours inside the selection box of Fig. \ref{colorsdiag}. In M and L dwarfs, sources with $\rm{i-z} \gtrsim \rm{i_{lim}-z_{lim}} = 0.9$ ($\rm{i_{lim} = 24.8}$ and $\rm{z_{lim} = 23.9}$ are the 80\% completeness limits for point sources in the i and z bands respectively, from the CFHTLS-T0007 release\footnote{\url{http://www.cfht.hawaii.edu/Science/CFHTLS/T0007/}}.) can be undetected in the i band, and therefore they are likely to be scattered into our candidate list. This limit is represented by a vertical dashed line in Fig. \ref{justifyBD}, where the observed colours of low-mass stars and of our z-band detected candidates are shown. \\
Here again, noise can scatter $\rm{z-Y}$ colours into our $\rm{z-Y}$ $\geq$ 1.5 criterion, and we therefore chose to adopt a more conservative limit for the contamination by low-mass stars of $\rm{z-Y} > 1.0$. Figure \ref{justifyBD} shows that this approximately corresponds to an L0 spectral type. For this reason, our brown dwarf model is based on a spectroscopic sample of 633 L dwarfs and 180 T dwarfs from the SpeX Prism library, with spectral types from L0 to T9. Hereafter, we assume that these observed spectra are representative of the colour distribution of the entire L- and T-dwarf populations.\\

We chose two parameters to model the brown dwarf population, the apparent magnitude $\rm{Y_{mod}}$ , and the spectral type spt: $\rm{\overrightarrow{p} = \left \{  Y_{mod}, spt\right \}}$. \\
We computed the brown dwarf surface density $\rm{\rho_{s}(\overrightarrow{p})}$ using the Galactic spatial density model of late-L and T dwarfs developed by \citet{Caballero2008}. With this model, we calculated the spatial density distribution of late-type dwarfs at a given Galactic coordinate as a function of luminosity for each spectral type. A complete description of the spatial variation of the density of late dwarfs is given in \citet{Caballero2008}, so we here report only the main calculation steps. We considered the standard Galactic model and followed the same approach as \citet{Caballero2008} by only taking into account the Galactic thin disc, which can be modelled by a double exponential law \citep[e.g.][]{Chen2001}. As discussed in detail in \citet{Caballero2008}, brown dwarfs from the Galactic thick disc and halo can be neglected for two reasons. First, the thick disc and the halo are more rarified and extended than the thin disc. Secondly, thick-disc and halo stars are thought to be very old \citep[ages $\gtrsim$ 10 Ga, see e.g.][]{Feltzing2003}: they are therefore extremely faint and hardly detectable, according to the general cooling theory for brown dwarfs \citep[see e.g.][]{Baraffe2003}. The space density for an object of spectral type $\rm{spt}$ at Galactic coordinates $(l,b)$ and heliocentric distance $d$ is given by

\begin{equation}
\label{eqtrue1}
n_{\rm{spt}}(d,l,b) = n_{0,\rm{spt}}\,\exp\left({ -\frac{R(d,l,b)-R_{\odot}}{h_{R}}}\right) \exp\left({-\frac{|Z_{\odot}+dsin(b)|}{h_{Z}}}\right)
\mathrm{,}
\end{equation}

where $Z_{\odot}$ is the height of the Sun relative to the Galactic plane, $h_{R}$ and $h_{Z}$ are the length and height scales of the thin disc, respectively, and $n_{0,\rm{spt}}$ is the space density at the Galactic plane ($Z=0$) and at the distance to the Galactic centre ($R = R_{\odot}$). The values of the main parameters of the Galaxy thin disc are listed in Table \ref{tablebd}.
        
Since CFHQSIR observations correspond to extragalactic fields, the Galactocentric distance of the object $R(d,l,b)$ can be approximated as follows, under the assumption $R_{\odot} \gg d$:

\begin{equation}
\label{eqapprox1}
R(d,l,b) \approx R_{\odot} - d\cos(b)\cos(l)
\mathrm{.}
\end{equation}

In this approximation, Eq. \ref{eqtrue1} can be rewritten as

\begin{equation}
\label{eqapprox2}
n_{\rm{spt}}(d,l,b) \approx n_{0,\rm{spt}}\, \exp \left({\frac{\mp Z_{\odot}}{h_{Z}}}\right) \exp \left[{-d\left(-\frac{\cos(b)\cos(l)}{h_{R}}\pm \frac{\sin(b)}{h_{Z}}\right)}\right]
\mathrm{,}
\end{equation}

where the sign convention indicates if $d\sin(b)$ is greater or lower than $Z_{\odot}$, that is, whether the object is above or below the Galactic plane. At null heliocentric distance, the local spatial density of an object of spectral type $\rm{spt}$, noted $\rm{n_{\odot}}$, is then

\begin{equation}
\label{denslocal}
n_{\rm{spt}}(R = R_{\odot}, Z = Z_{\odot}) = n_{0,\rm{spt}}\, \exp\left({-\frac{Z_{\odot}}{h_{Z}}}\right) = n_{\odot}
\mathrm{.}
\end{equation}

\begin{figure}\centering
	\includegraphics[width=0.45\textwidth]{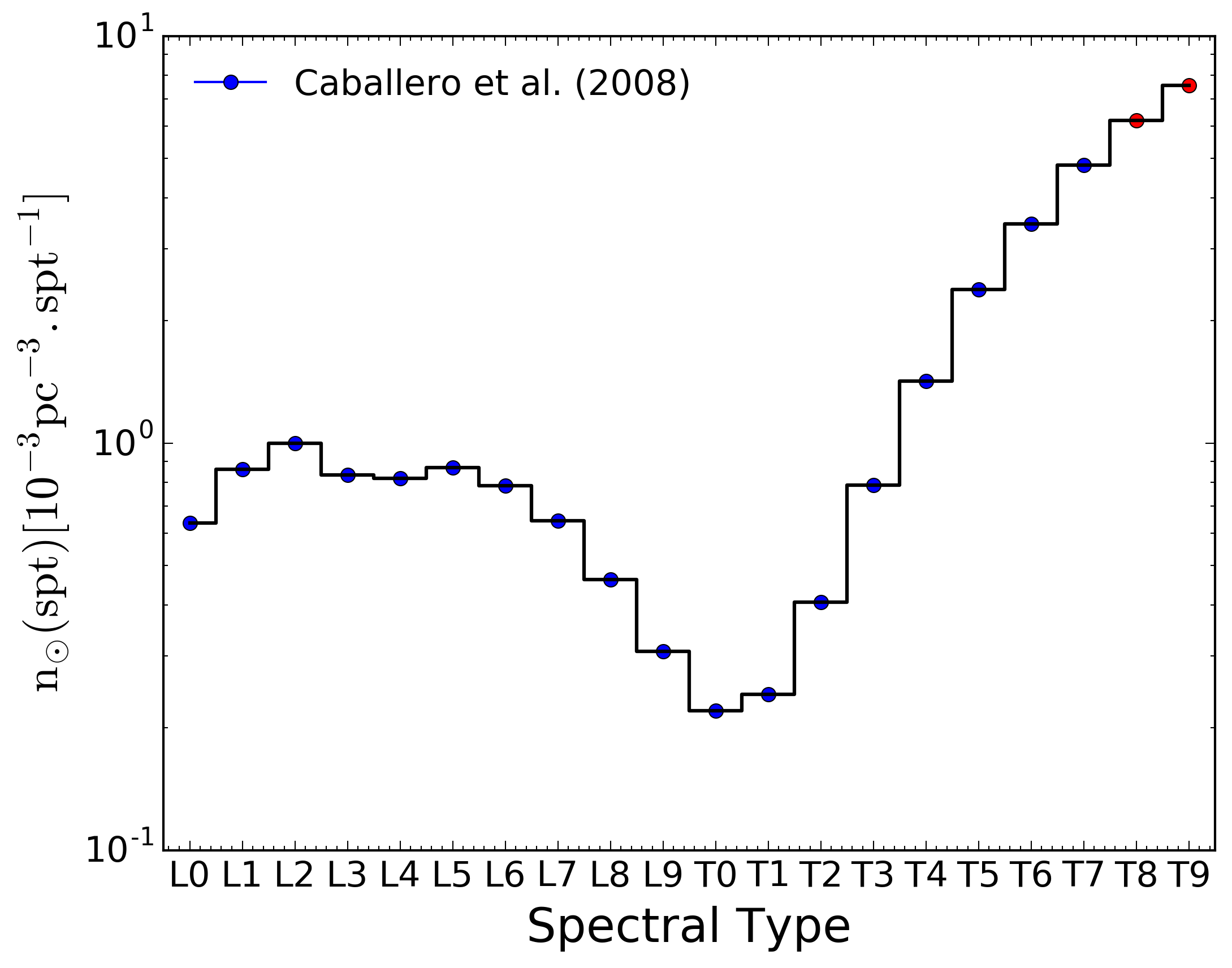}
	\caption{Local spatial density $\rm{n_{\odot}}$ of late dwarfs as a function of spectral type. The blue points refer to the values given in \citet{Caballero2008}, and the red points correspond to extrapolated values.}
	\label{nsol}
\end{figure}

Fig. \ref{nsol} shows $\rm{n_{\odot}}$ as a function of spectral type. For L0-T7 dwarfs, we used the predicted densities reported in \citet{Caballero2008} that we took from \citet{Burgasser2007}. The densities for the T8 and T9 spectral types are extrapolated values.\\
        
\begin{table}[ht]
	\caption{Main parameters of the Galactic thin disc used in \citet{Caballero2008} adopted from \citet{Chen2001}}              
	\label{tablebd}      
	\centering                                      
	\begin{tabular}{c c c c}          
		\hline\hline                        
		$R_{\odot}$(pc) & $Z_{\odot}$(pc) & $h_{R}$(pc) & $h_{Z}$(pc) \\    
		\hline                                   
		$8600\pm200$ & $+27\pm4$ & $2250\pm1000$ & $330\pm3$ \\      
		\hline                                             
	\end{tabular}
\end{table}    
        
\citet{Caballero2008} defined two auxiliar variables, $n_{A,\rm{spt}}$ and $d_{B}(l,b)$ to simplify calculations:

\begin{equation}
n_{A,\rm{spt}} = n_{0,\rm{spt}}\,\exp\left({\frac{\mp Z_{\odot}}{h_{Z}}}\right) = \begin{cases}
n_{A+,\rm{spt}} = n_{0,\rm{spt}}\,\exp \left({\frac{- Z_{\odot}}{h_{Z}}}\right)\,\,\text{if $Z > 0$} \\ \\
n_{A-,\rm{spt}} = n_{0,\rm{spt}}\,\exp \left({\frac{+ Z_{\odot}}{h_{Z}}}\right)\,\,\text{if $Z < 0$} 
\end{cases}
\mathrm{,}
\end{equation}

\begin{equation}
\label{db}
\frac{1}{d_{B}(l,b)} = -\frac{\cos(b)\cos(l)}{h_{R}} \pm \frac{\sin(b)}{h_{Z}} 
\mathrm{,}
\end{equation}

with

\begin{equation}
\label{db2}
\begin{cases}
\frac{1}{d_{B+}(l,b)} = -\frac{\cos(b)\cos(l)}{h_{R}} + \frac{\sin(b)}{h_{Z}} \,\,\text{if $Z > 0$},\\ \\
\frac{1}{d_{B-}(l,b)} = -\frac{\cos(b)\cos(l)}{h_{R}} - \frac{\sin(b)}{h_{Z}} \,\,\text{if $Z < 0$}, 
\end{cases}
\end{equation}

Based on Eqs. 12 and 14 of \citet{Caballero2008}, we computed the brown dwarf surface density per spectral type $\rm{\rho_{s}(\textit{d},\rm{spt})}$ (in sr$^{-1}\,$spt$^{-1}$) as a function of the maximum heliocentric distance $d$ with

\begin{equation}
\label{rhoint}
\rho_{s}(d,\rm{spt}) = \begin{cases}
n_{A+,\rm{spt}} \int_{0}^{d} \exp \left({-\frac{z}{d_{B+}}}\right) z^{2}dz\,\,\text{if the field $\in$ NGP,}\\ \\
n_{A+,\rm{spt}} \int_{0}^{d_{*}} \exp \left({-\frac{z}{d_{B+}}}\right) z^{2}dz \\+ \, n_{A-,\rm{spt}} \int_{d_{*}}^{d} \exp \left({-\frac{z}{d_{B-}}}\right) z^{2}dz\,\,\text{if the field $\in$ SGP,}\\
\end{cases}
\end{equation}

where $d_{*} = -Z_{\odot}/\sin(b)$, and NGP and SGP refer to the north and south Galactic poles, respectively. We finally computed the maximum heliocentric distance $d$ beyond which a source with an absolute magnitude $\rm{M_{Y_{mod}}}$ can be detected, by solving the following equation: 

\begin{equation}
\rm{Y_{mod} - M_{Y_{mod}} = 5log(\textit{d}) - 5 + A_{Y_{mod}}}
\mathrm{,}
\label{Ymoddens}
\end{equation}

where $\rm{A_{Y_{mod}}}$ corresponds to the attenuation in Y band. Table \ref{tab} presents an example of the obtained values for the distance $d$, considering a brown dwarf with an apparent magnitude $\rm{Y_{mod}}$ = 22.5.\\
Using Eq. \ref{rhoint} and \ref{Ymoddens}, we estimated the brown dwarf surface density $\rm{\rho_{s}(Y_{mod},spt)}$ as a function of the Y-band apparent magnitude for each spectral type in \mbox{mag$^{-1}$ deg$^{-2}$ spt$^{-1}$} units.

\begin{table*}
        \caption{Characteristics of L and T dwarfs}
        \centering 
        \begin{tabular}{ccccccc}   
                \hline          \hline
                \multirow{2}{*}{
                        \centering Spectral type spt} &
                \multirow{2}{*}{
                        \centering $\rm{M_{Y_{mod}}}$ (mag)} & 
                \multirow{2}{*}{
                        \centering $\rm{n_{\odot}}$ (\mbox{10$^{-3}$ pc$^{-3}$ spt$^{-1}$})} & 
                
                \multicolumn{4}{c}{\multirow{2}{*}{$d_{\rm{Y_{mod}}=22.5}$ (pc)}} \\ \\
                
                &  &  & W1 & W2 & W3 & W4 \\
                \hline
                \\
                L0 & 13.21  & 0.637  & 695 & 702 & 687 & 666 \\
                L1 & 13.62  & 0.861  & 579 & 584 & 573 & 558 \\
                L2 & 14.09  & 1.000  & 470 & 473 & 466 & 456 \\
                L3 & 14.54  & 0.834  & 383 & 385 & 380 & 374 \\
                L4 & 15.00  & 0.819  & 311 & 312 & 309 & 305 \\
                L5 & 15.40  & 0.869  & 259 & 260 & 258 & 255 \\
                L6 & 15.76  & 0.785  & 220 & 221 & 220 & 217 \\
                L7 & 16.09  & 0.644  & 190 & 191 & 190 & 188 \\
                L8 & 16.21  & 0.462  & 180 & 180 & 179 & 178 \\
                L9 & 16.36  & 0.308  & 167 & 168 & 167 & 166 \\
                T0 & 16.39  & 0.22  & 165 & 166 & 165 & 163 \\
                T1 & 16.39  & 0.241  & 166 & 166 & 165 & 164 \\
                T2 & 16.36  & 0.406  & 167 & 168 & 167 & 165 \\
                T3 & 16.23 & 0.788 & 178 & 178 & 177 & 176 \\
                T4 & 16.32  & 1.42  & 171 & 171 & 170 & 169 \\
                T5 & 16.35  & 2.38  & 168 & 169 & 168 & 166 \\
                T6 & 16.60  & 3.45  & 150 & 151 & 150 & 149 \\
                T7 & 17.11  & 4.82  & 119 & 119 & 119 & 118 \\
                T8 & 17.96  & 6.19  & 81 & 81 & 81 & 81 \\
                T9 & 19.25  & 7.56  & 45 & 45 & 45 & 44 \\
                \hline
        \end{tabular}
        \tablefoot{$d$ is the maximum heliocentric distance for which a brown dwarf of a given spectral type and with an apparent magnitude $\rm{Y_{mod} = 22.5}$ can be detected. The absolute magnitudes $\rm{M_{Y_{mod}}}$ were computed by combining the absolute magnitude in the 2MASS J filter $\rm{M_{J_{2MASS}}}$ derived by \citet{Dupuy2012} for each spectral type, with the observed colours $\rm{Y_{mod} - J_{2MASS}}$ measured on the 813 brown dwarf spectra used in this study. We also reference the local spatial density $\rm{n_{\odot}}$ .} 
        \label{tab}
\end{table*}

The weighted evidence for the brown dwarf population in \mbox{mag$^{-3}$ deg$^{-2}$} is given by

\begin{equation}
\label{poidsbd}
\begin{multlined}
W_{s}(\rm{z_{obs}},\rm{Y_{obs}},\rm{J_{obs}},\rm{det}) \\
= \int_{L0}^{T9} \int_{-\infty}^{+\infty} \rho_{s}(\rm{Y_{mod}},\rm{spt})\,\rm{Pr(det|Y_{mod},spt)}\\
\times \rm{Pr(z_{obs}, Y_{obs}, J_{obs}| Y_{mod},spt)}\,\rm{dY_{mod}}\,\rm{dspt} ,
\end{multlined}
\end{equation}

where $\rho_{s}(\rm{Y_{mod}},\rm{spt})$ is the surface density previously estimated, and the probabilities $\rm{Pr(det|Y_{mod},spt)}$ and $\rm{Pr(z_{obs}, Y_{obs}, J_{obs}| Y_{mod},spt)}$ are defined in the same way as for the quasar population.\\
We note with $\rm{N_{s, spt}}$ the number of brown dwarf spectra available for a spectral type spt. We decided not to reduce one spectral type to a unique pair of colour ($\rm{z_{mod}-Y_{mod}}; \rm{Y_{mod}-J_{mod}})$ but to consider instead each brown dwarf spectrum individually. This allowed us to estimate and account for the dispersion within a single spectral type. We first computed the weighted evidence associated with the $i$-th brown dwarf of spectral type spt, in \mbox{mag$^{-3}$ deg$^{-2}$ spt$^{-1}$}: 

\begin{equation}
\begin{multlined}
\rm{w}_{s,spt,i}(\rm{z_{obs}},\rm{Y_{obs}},\rm{J_{obs}},\rm{det}) \\
= \int_{-\infty}^{+\infty} \rho_{s}(Y_{mod,i},spt)\,\rm{Pr(det|Y_{mod,i},spt)}\\
\times \rm{Pr(z_{obs}, Y_{obs}, J_{obs}| Y_{mod,i},spt)}\,\rm{dY_{mod,i}}.
\end{multlined}
\end{equation}

The probability $\rm{Pr(z_{obs}, Y_{obs}, J_{obs}| Y_{mod,i},spt)}$ was calculated in the same way as for the high-z quasar population: using Eq. \ref{vraisemblance} and associating the modeled flux with the observed brown dwarf colours. Each brown dwarf $i$ contributes the same to the weighted evidence associated with the spectral type spt, so that we can write 

\begin{equation}
\label{moyspt}
\rm{w}_{\rm{s,spt}}(\rm{z_{obs}},\rm{Y_{obs}},\rm{J_{obs}},\rm{det})= \frac{1}{\rm{N_{s,spt}}} \sum_{i=1}^{\rm{N_{s,spt}}} w_{\rm{s,spt},i}(\rm{z_{obs},Y_{obs},J_{obs},det}).
\end{equation}

Eq. \ref{poidsbd} then becomes

\begin{equation}
\rm{W_{s}}(\rm{z_{obs},Y_{obs},J_{obs},det}) = \int_{L0}^{T9} w_{\rm{s,spt}}(\rm{z_{obs},Y_{obs},J_{obs},det})\,\rm{dspt}, 
\end{equation}

where dspt = 1 spt, with $\rm{spt}(L0) = 0$ and $\rm{spt}(L1) = 1$ etc. are as defined in \citet{Burgasser2007}. The latter integral can therefore be reduced to a simple sum over spectral types L0 to T9. \\

We based our Bayesian classification method on the formalism proposed by \citet{Mortlock2012} and improved the quasar and low-mass star modelling in several aspects that we describe below. \citet{Mortlock2012} used 12 different quasar templates spanning four line-strengths and three continuum slopes to account for the intrinsic diversity of the quasar population, while we based our high-z quasar model on 80 simulated quasar spectra, each with different spectral indices, emission line widths, and intensities. Furthermore, we developed an enhanced brown dwarf model that better reflects their properties. We included the variation in space density with Galactic coordinates, while \citet{Mortlock2012} adopted a model corresponding to an average of the brown dwarf population over the range of Galactic latitudes covered by the UKIDSS LAS. Finally, we included the intrinsic colour spread in $\rm{Y-J}$ for L and T dwarfs, whereas \citet{Mortlock2012} attributed only one $\rm{Y-J}$ colour to each spectral type. \\
Our Bayesian probabilistic algorithm is similar to the model developed by \citet{Matsuoka2016} for the SHELLQs project, but has two notable differences. First, \citet{Matsuoka2016} needed to include M dwarfs in their model in addition to L and T dwarfs since they probed a lower redshift range ($5.7 < z < 6.9$). Secondly, they used a unique quasar spectral energy distribution (SED) from the stacking of 340 bright SDSS quasar spectra at $z \sim 3$, while our model uses 80 different spectra created from a principal component analysis of 78 SDSS spectra at $z \sim 3$ (see Sects. \ref{simulqso} and \ref{qsopop}).

\subsubsection{Probability of being a high-z quasar: simulations}

\begin{figure*}[ht!]
        \centering
        \includegraphics[width=0.85\textwidth]{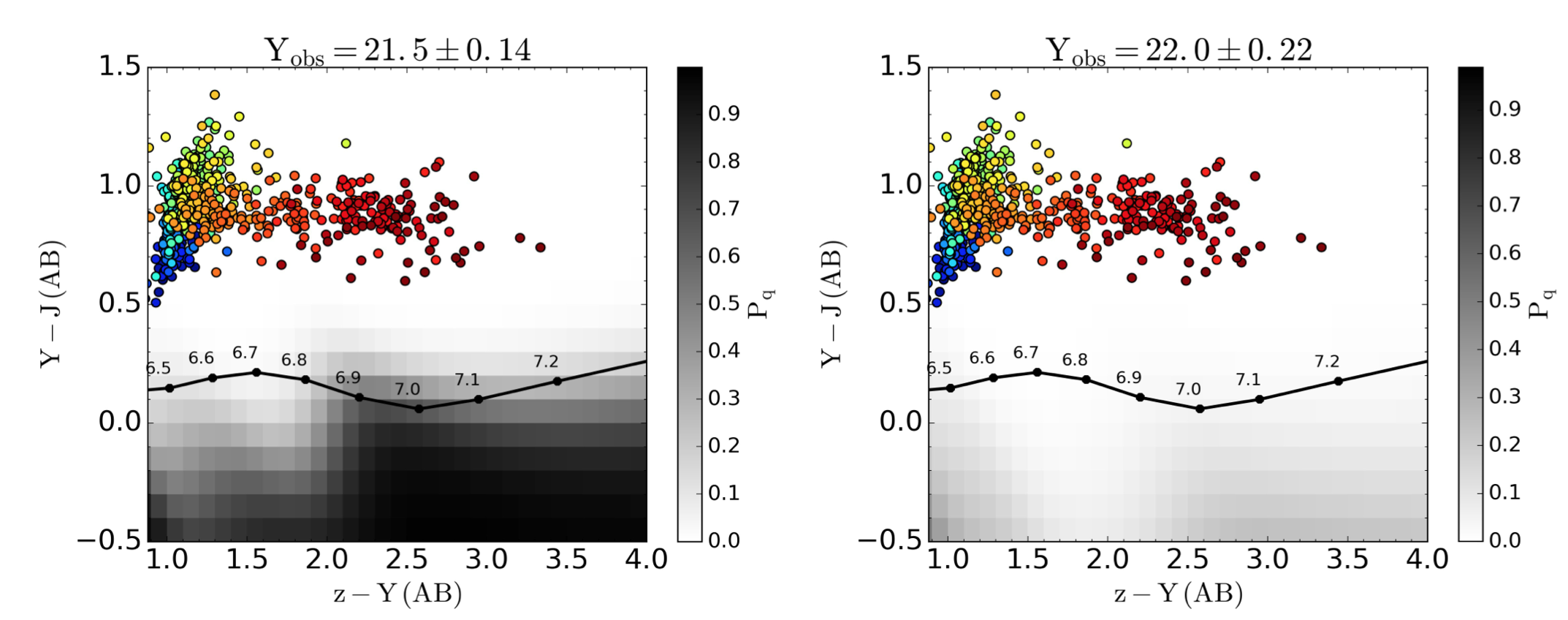}
        \caption{Probability $\rm{P_{q}}$ of being a high-redshift quasar for simulated sources with an observed magnitude $\rm{Y_{obs} = 21.5 \pm 0.14}$ (left panel) and $\rm{Y_{obs} = 22.0 \pm 0.22}$ (right panel). This probability reaches from white ($\rm{P_{q}} = 0$) to black ($\rm{P_{q}} = 1$). The coloured points represent brown dwarf colours ranging from blue (L0) to red (T9). The black curve is the mean quasar spectrum derived by \citet{Paris_2011}, for which we included IGM absorption.}
        \label{Pqsimulmap}
\end{figure*}

\begin{figure}\centering
        \includegraphics[width=0.35\textwidth]{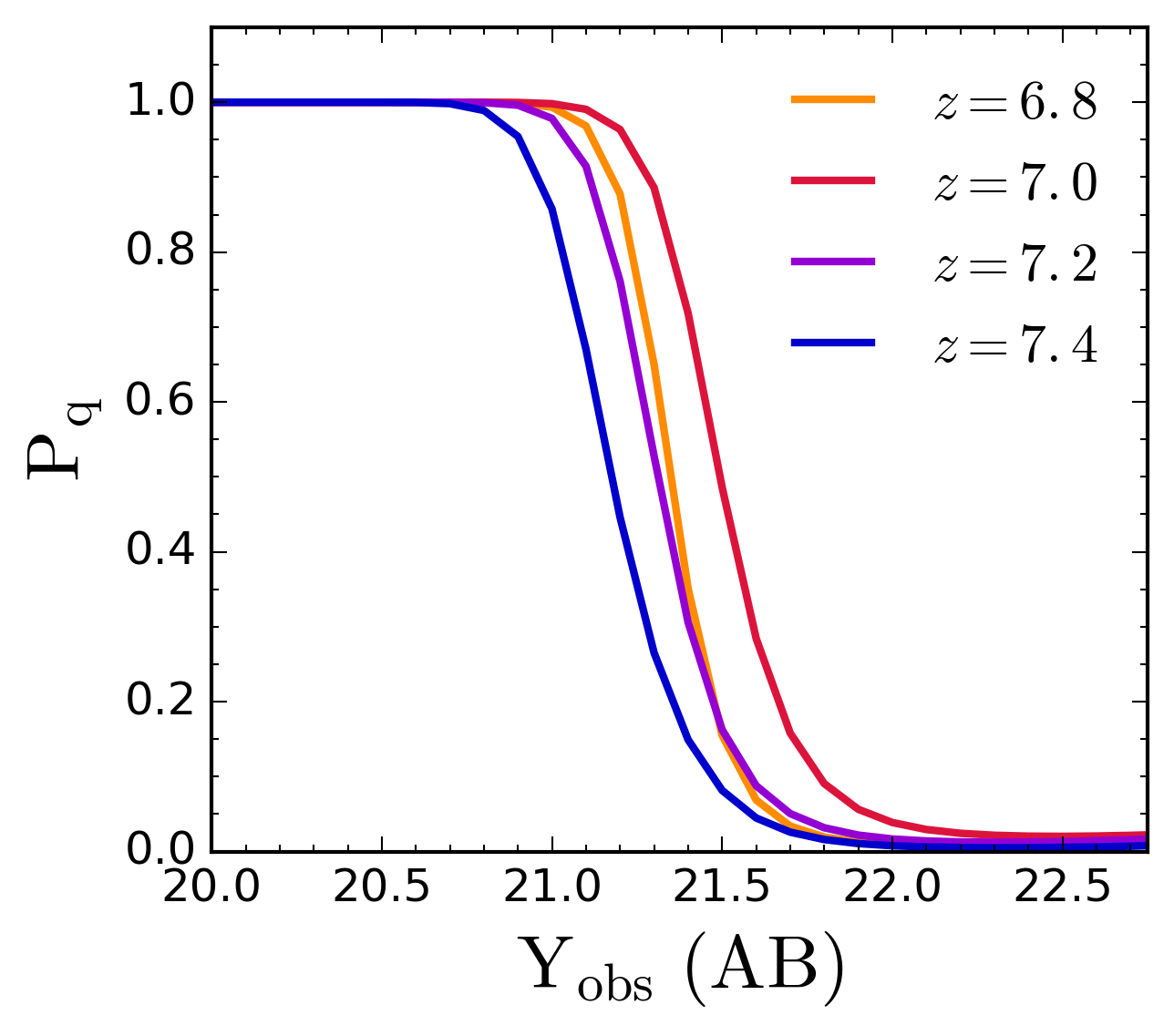}
        \caption{Probability $\rm{P_{q}}$ for a source observed with the $\rm{(z_{obs}-Y_{obs})}$ and $\rm{(Y_{obs}-J_{obs})}$ colours of a z = $\left \{6.8, 7.0, 7.2, 7.4\right \}$ quasar as a function of its measured Y-band magnitude $\rm{Y_{obs}}$.}
        \label{PqsimulcourbeY}
\end{figure}

In this section, we explore the dependence of the probability $\rm{P_{q}}$ on photometric measurements. For this purpose, we simulated a set of sources uniformly distributed in the $\rm{z_{obs}-Y_{obs}}$ versus $\rm{Y_{obs}-J_{obs}}$ diagram with CFHQSIR-like z-, Y-, and J-band measurements. Figure \ref{Pqsimulmap} shows the probability $\rm{P_{q}}$ obtained for sources observed with an apparent magnitude $\rm{Y_{obs}} = 21.5 \pm 0.14$ (left panel) and $\rm{Y_{obs}} = 22.0 \pm 0.22$ (right panel). As expected, $\rm{P_{q}}$ increases when the object approaches the characteristic high-z quasar colours. The comparison between the two panels clearly shows that $\rm{P_{q}}$ strongly depends on the observed magnitude $\rm{Y_{obs}}$ , and its high-probability region (in black) is much more prononced for the brightest objects (left panel). This effect is mostly related to the photometric errors: bright objects are measured with a sufficiently high S/N for there to be no ambiguity about their nature. However, faint objects are more likely to be stars measured with high-z quasar colours because of a lower S/N and greater number of stars compared to high-z quasars.\\
This dependency is also demonstrated in Fig. \ref{PqsimulcourbeY}, where the probability $\rm{P_{q}}$ for sources with quasar colours simulated from the mean quasar spectrum discussed in Sect. \ref{simulqso} at four different redshifts is represented as a function of their observed Y-band magnitude. Well-detected sources have a probability $\rm{P_{q}} = 1$, and their nature is unambiguously established. However, for fainter objects, $\rm{P_{q}}$ drops sharply. As mentioned above for the previous figure, the fact that brown dwarfs outnumber the quasars combined with lower S/N implies that these sources are more easily construed as brown dwarfs scattered into the quasar loci. The redshift dependency lies in the slight variation of the $\rm{Y_{mod} - J_{mod}}$ quasar colours. As shown in Fig. \ref{colorsdiag},  the $\rm{Y_{mod} - J_{mod}}$ colour of a $z = 7.2$
quasar is much closer to the brown dwarf locus (with $\rm{Y_{mod} - J_{mod}} \approx 0.3$) than that of a $z = 7.0$ quasar, for which $\rm{Y_{mod} - J_{mod}} \approx 0.1$ because the Ly-$\alpha$ line enters the Y band.\\
In the next section, we present our high-z quasar candidate probabilities and discuss our results. 

\begin{figure}[h!]\centering
        \includegraphics[width=0.45\textwidth]{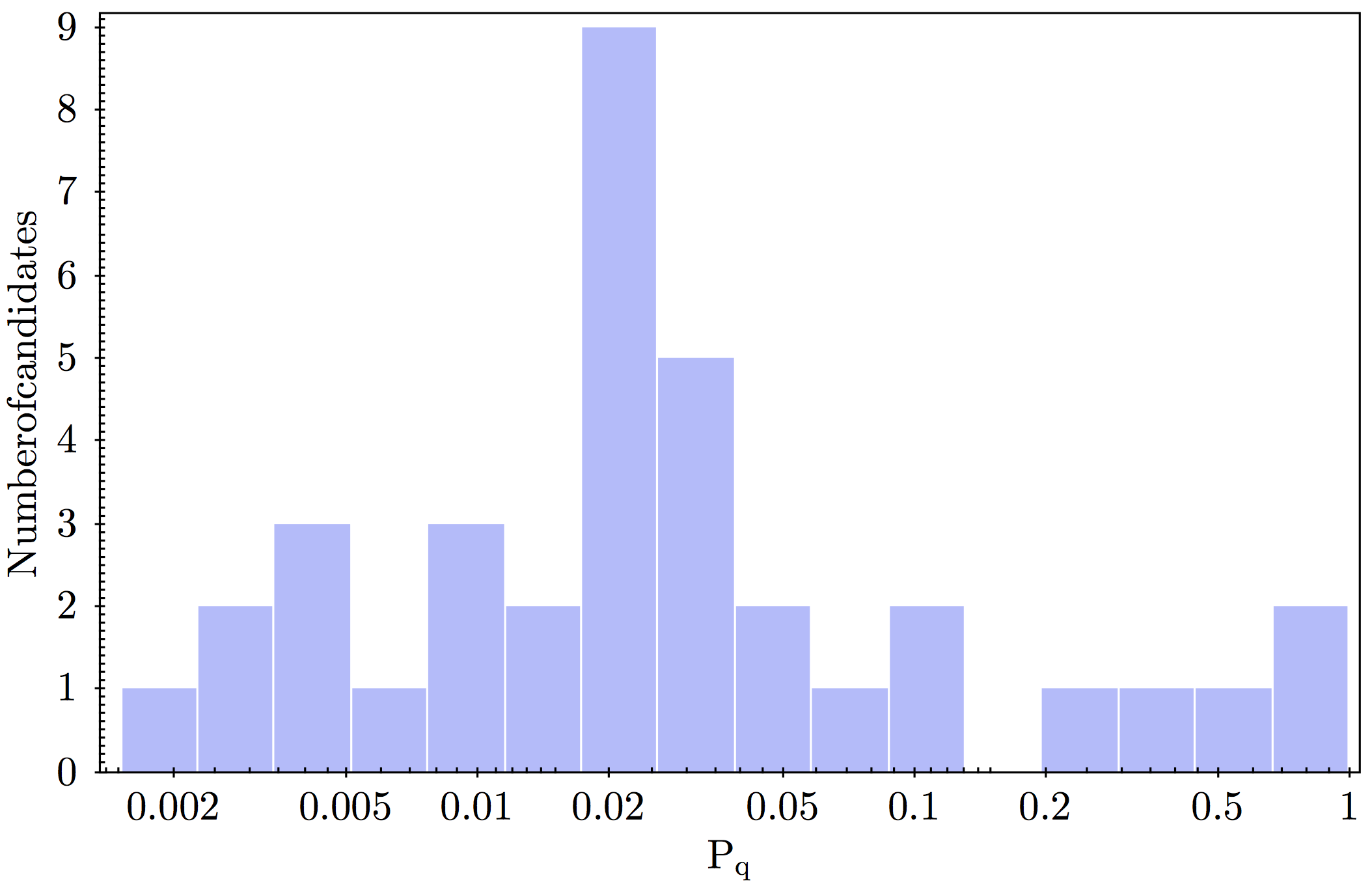}
        \caption{Histogram of the high-z quasar candidate probabilities. Fewer than 20\% of our colour-selected candidates have a probability $\rm{P_{q} > 0.1}$.}
        \label{Pqhist}
\end{figure}

\section{Results}
\label{secresults}
\begin{figure*}[ht!]
        \centering
        \includegraphics[width=0.85\textwidth]{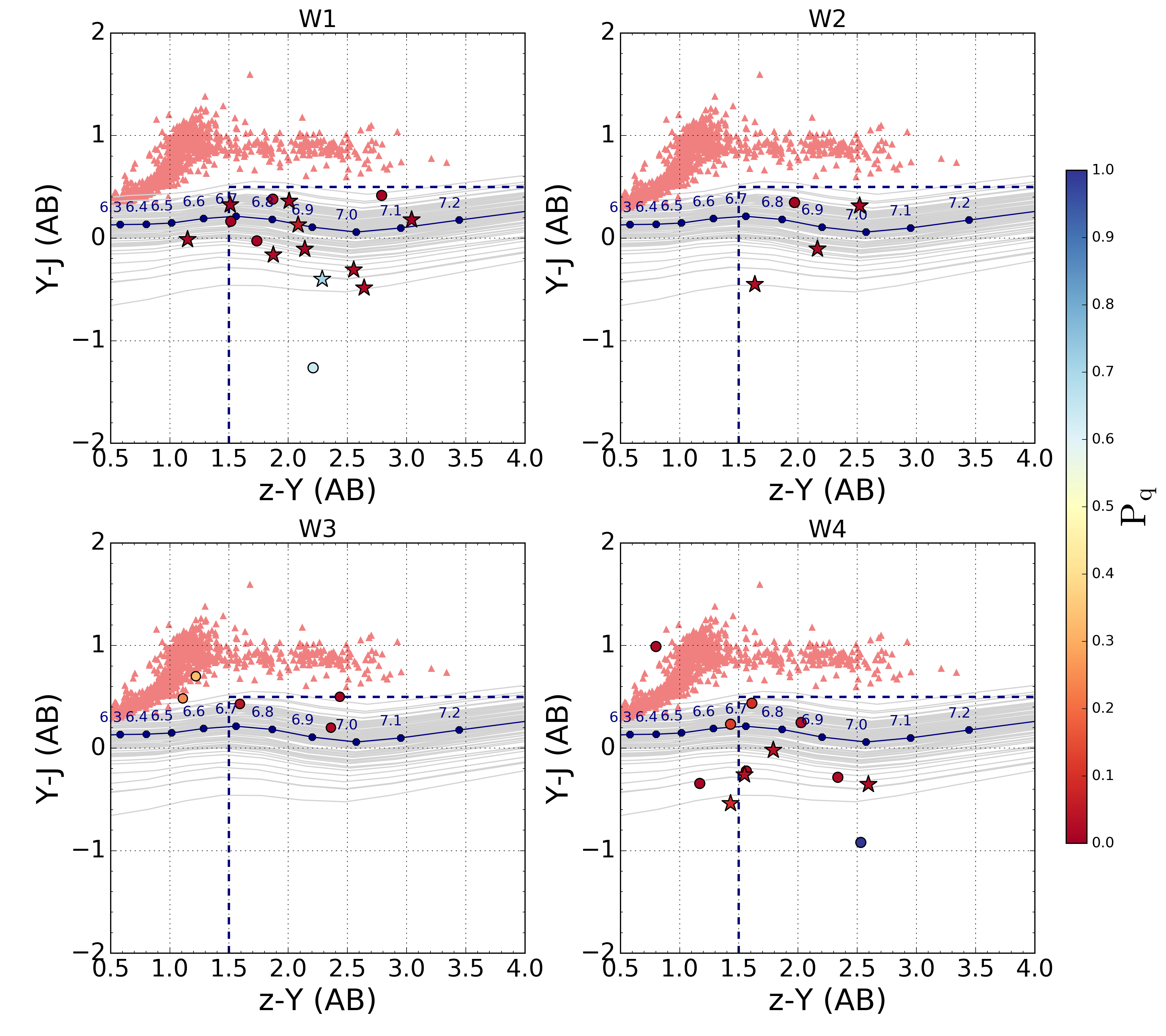}
        \caption{Colour-colour diagram that represents our high-z quasar candidates according to a colour-code from red to blue, corresponding to low and high probability $\rm{P_{q}}$. Stars indicate spectroscopically observed candidates (NTT or VLT).}
        \label{Pqcands}
\end{figure*}

We applied the Bayesian formalism developed in the previous section to our list of 51 high-z quasar candidates. We calculate the $\rm{P_{q}}$ value for each candidate and report the resulting distribution in Fig. \ref{Pqhist}. As expected, the main result is that most of our candidates have a low probability ($\rm{P_{q}} \approx 10^{-2}$). Only 6 of 36 candidates have a chance greater than 10\%  ($\rm{P_{q}} > 0.1$) of being high-z quasars. Three of them have a probability $\rm{P_{q}} > 0.6$. This result arises because high-z quasars are outnumbered by low-mass stars and the vast majority of our candidates has a relatively faint Y-band magnitude ($\rm{Y_{obs} \sim 22.4}$ on average). \\
Figure \ref{Pqcands} corresponds to the same colour-colour diagrams as shown in Fig. \ref{diag_cands}, but with the difference that the probabilities $\rm{P_{q}}$ are indicated according to a colour-code from red (low probability) to blue (high probability). Stars refer to candidates observed spectroscopically. In total, 17 candidates have been followed-up in spectroscopy in the W1, W2, and W4 fields using the infrared spectrograph SOFI at NTT (GBF grism blue) and the optical spectrograph FORS2 at the Very Large Telescope (VLT; GRIS\_600z+23 grism). Positions, photometry, probabilities, and spectroscopic informations of these sources are given in Table \ref{table:1}. Of these, 14 with $\rm{P_{q}} \lesssim 0.2$ were consistent with being late brown dwarfs. The remaining 3 candidates observed with FORS2 at the VLT need to have further spectroscopic observations in the infrared because their spectral S/N is too low to confidently confirm their exact nature at this moment. However, as shown in Fig. \ref{Pqcands}, some candidates with a high probability have not yet been observed spectroscopically (one candidate in the W1 field with $\rm{P_{q}} \sim 0.6$ and another in the W4 field with $\rm{P_{q}} \sim 1.0$) because they were judged to be too faint in the z band to be detected in optical spectroscopy with FORS2. 

\begin{sidewaystable*}
        \caption{Positions, photometry, and probabilities of being a high-z quasar and spectroscopic informations of the spectroscopically observed quasar candidates}              
        \label{table:1}      
        \centering       
        \begin{tabular}{c c c c c c c}          
                \hline\hline                        
            Field, RA and DEC (J2000.0) & Y (mag) & $\rm{z-Y}$ & $\rm{Y-J}$ & $\rm{P_{q}}$ & Telescope/Instrument & Grism\\    
                \hline                                   
                W1 02:06:24.64 -07:03:50.73 & $\rm{22.91 \pm 0.56}$ & $\rm{1.87 \pm 0.74}$ & $\rm{-0.16 \pm 0.63}$ & $2.4\times 10^{-2}$ & VLT/FORS2 & GRIS\_600z + 23\\      
                W1 02:07:25.13 -05:29:25.00 & $\rm{22.74 \pm 0.40}$ & $\rm{2.55 \pm 0.75}$ & $\rm{-0.31 \pm 0.50}$ & $2.3\times 10^{-2}$ & VLT/FORS2 & GRIS\_600z + 23\\      
                W1 02:31:08.79 -05:01:12.80 & $\rm{22.15 \pm 0.28}$ & $\rm{2.14 \pm 0.44}$ & $\rm{-0.11 \pm 0.31}$ & $1.2\times 10^{-2}$ & VLT/FORS2 & GRIS\_600z + 23 \\      
                W1 02:06:50.85 -10:09:33.47 & $\rm{22.12 \pm 0.34}$ & $\rm{3.04 \pm 0.44}$ & $\rm{0.18 \pm 0.46}$ & $3.7\times 10^{-3}$ & VLT/FORS2 & GRIS\_600z + 23 \\      
                W1 02:34:05.68 -06:21:09.76 & $\rm{22.78 \pm 0.46}$ & $\rm{2.01 \pm 0.50}$ & $\rm{0.36 \pm 0.50}$ & $4.7\times 10^{-3}$ & VLT/FORS2 & GRIS\_600z + 23 \\      
                W1 02:25:59.13 -10:34:50.71 & $\rm{22.45 \pm 0.32}$ & $\rm{2.64 \pm 0.39}$ & $\rm{-0.49 \pm 0.45}$ & $2.3\times 10^{-2}$ & VLT/FORS2 & GRIS\_600z + 23\\      
                W1 02:09:47.18 -08:23:52.10& $\rm{22.48 \pm 0.34}$ & $\rm{>1.51}$ & $\rm{0.33 \pm 0.47}$ & $2.7\times 10^{-3}$ & NTT/SOFI & GB Grism Blue\\      
                W1 02:26:56.80 -04:24:06.88 & $\rm{22.20 \pm 0.24}$ & $\rm{2.29 \pm 0.34}$ & $\rm{<-0.40}$ & $7.1\times 10^{-1}$ & VLT/FORS2 & GRIS\_600z + 23\\      
                W1 02:07:57.65 -08:04:26.00 & $\rm{22.67 \pm 0.38}$ & $\rm{2.08 \pm 0.60}$ & $\rm{<0.13}$ & $5.6\times 10^{-2}$ & VLT/FORS2 & GRIS\_600z + 23\\      
                W1 02:17:29.38 -05:01:31.49 & $\rm{22.81 \pm 0.41}$ & $\rm{>1.15}$ & $\rm{-0.01 \pm 0.66}$ & $1.9\times 10^{-2}$  & VLT/FORS2 & GRIS\_600z + 23\\      
                W2 09:04:32.26 -05:31:56.26 & $\rm{22.38 \pm 0.27}$ & $\rm{1.64 \pm 0.35}$ & $\rm{-0.45 \pm 0.39}$ & $3.0\times 10^{-2}$ & VLT/FORS2 & GRIS\_600z + 23\\      
                W2 09:04:53.35 -03:15:22.06 & $\rm{22.77 \pm 0.36}$ & $\rm{2.16 \pm 0.49}$ & $\rm{-0.11 \pm 0.45}$ & $2.1\times 10^{-2}$ & VLT/FORS2 & GRIS\_600z + 23\\      
                W2 09:04:22.51 -04:34:09.71 & $\rm{21.84 \pm 0.19}$ & $\rm{>2.52}$ & $\rm{0.32 \pm 0.23}$ & $8.4\times 10^{-3}$  & NTT/SOFI & GB Grism Blue\\      
                W4 22:17:59.64 00:55:02.12 & $\rm{22.85 \pm 0.46}$ & $\rm{1.79 \pm 0.62}$ & $\rm{-0.02 \pm 0.49}$ & $3.9\times 10^{-2}$ & VLT/FORS2 & GRIS\_600z + 23\\      
                W4 22:02:41.32 01:34:31.95 & $\rm{22.44 \pm 0.25}$ & $\rm{1.55 \pm 0.34}$ & $\rm{-0.26 \pm 0.33}$ & $3.2\times 10^{-2}$ & VLT/FORS2 & GRIS\_600z + 23\\      
                W4 22:04:12.48 02:01:23.22& $\rm{22.61 \pm 0.39}$ & $\rm{2.59 \pm 0.59}$ & $\rm{-0.35 \pm 0.43}$ & $2.7\times 10^{-2}$ & VLT/FORS2 & GRIS\_600z + 23\\      
                W4 22:10:33.26 +02:20:24.89& $\rm{22.38 \pm 0.27}$ & $\rm{>1.43}$ & $\rm{<-0.54}$ & $8.6\times 10^{-2}$ & NTT/SOFI & GB Grism Blue\\      

                \hline                                             
        \end{tabular}
\tablefoot{No quasar was confirmed after these spectroscopic observations. The detailed analysis of the spectroscopic results will be reported in a future paper.}
\end{sidewaystable*}

\section{Discussion and conclusions}
\label{sec5}

We have presented a complete method for selecting and classifying $z \sim 7$ quasar candidates using optical and NIR photometric data as part of the CFHQSIR survey. After visual inspection and removal of artefacts, we selected 228 candidates with red $\rm{z-Y}$ colours. The sample was eventually culled to 36 sources by considering candidates with $\rm{Y-J}$ colours consistent with being high-z quasars after NIR follow-up in J band. We extended and refined the Bayesian formalism developed by \citet{Mortlock2012} in order to improve our selection of high-z quasar candidates. This robust statistical approach allowed us to classify our candidates in the best possible way according to their probability of being a high-redshift quasar. Applying a Bayesian method is indeed very efficient to identify the most promising candidates in any photometrically selected sample since it allows combining the prior knowledge of the brown dwarf and quasar populations and the filter and noise properties of the observational data. Moreover, we demonstrated the importance of acquiring sufficiently precise photometric measurements to clearly determine the nature of the objects. \\
The application of the Bayesian formalism revealed a promising list of six candidates with a probability $\rm{P_{q}} > 0.1$. Only three of these have a chance higher than 60\% of being a high-z quasar. Even though one of them has been observed spectroscopically with the FORS2 instrument, we were unable to draw any conclusions about its nature because its spectral S/N is low. Our immediate goal is to complete the spectroscopic follow-up of our most promising candidates, according to their probability of being a high-z quasar. The full analysis of our search for $z \sim 7$ quasars will be reported in a future paper, including the analysis of our spectroscopic data. A non-detection indicates that the quasar density at $z \sim 7$ is at a 90\% and 75\% confidence level lower than the density inferred from the QLF of \citet{Willott2010} and  \citet{Jiang2016}, respectively. Conversely, the discovery of one high-redshift quasar in CFHQSIR would be 23\% and 35\% consistent with the QLF of \citet{Willott2010} and \citet{Jiang2016}, respectively.
\\

Our Bayesian method can be consolidated in several aspects, however. Early-type galaxies at $z \sim 2$ may appear compact at faint magnitudes and therefore represent a second source of contamination that would also need to be modelled precisely, in addition to the low-mass stars. \\
Furthermore, the brown dwarf population could be modelled more accurately by considering the stars from the Galactic thick disc and the halo, in addition to the thin disc. There is also further room for improvement concerning the local spatial density of brown dwarfs. Using observed spatial density measurements instead of model-predicted ones would be interesting \citep[see e.g.][]{Reyle2010, Kirkpatrick2011}. As the number of known ultracool dwarfs continues to rise, more precise estimates of their spatial densities will be essential ingredients of an improved Bayesian model. Our results may also be biased as they are based on a heterogeneous collection of L- and T-dwarf spectra coming from various surveys, provided by the SpeX Prism library. In particular, our fundamental assumption that the measured colours of these sources are representative of the colours of the L and T populations may no longer be valid when considering unresolved binaries and peculiar sources. The unusual colours of these rare sources, as well as the uncertainty in their spectral classification, can introduce an intrinsic scatter in the colours that is not taken into account in our model. Unresolved binaries can be an important source of uncertainty, as they could represent $\sim$ 40\%  of the dwarf population \citep{Liu2006, Burgasser2007}. To finish, the newly discovered spectral type, the ultracool Y-dwarf class \citep[e.g. ][]{Cushing2011}, should also be included in our model, although these sources are likely to be too faint at our limiting magnitude to represent a real source of contamination.\\
Nevertheless, the Bayesian technique presented in this paper remains a powerful approach to efficiently prioritize high-z quasar candidates, and it is moreover easily adaptable to any other high-z object surveys, whether they are dedicated to the search for quasars or not (e.g. high-z galaxies in deep fields). This technique will be critical to the discovery of many more distant quasars by future NIR surveys such as the Euclid-wide imaging survey \citep{Laureijs2011} or the WFIRST High Latitude Survey \citep[HLS, ][]{Spergel2013}. Thanks to the high sensitivity of these upcoming surveys, high-z quasars will be identified in an unprecedented number. Scheduled to be launched in 2021 and in the mid-2020s, respectively the Euclid space mission is expected to discover $\sim$ 30 quasars at redshifts $z > 8.1$ with $J < 22.0$ over 18 000 $\rm{deg^{2}}$ , and WFIRST will find $\sim$ 500 $z \gtrsim 8$ quasars over 2\,000 $\rm{deg^{2}}$ above 10$\,\sigma$ \citep[mag $\approx$ 26 limit, ][]{Spergel2013}. \\
A great challenge for these high-depth surveys will be the increasing number of contamination sources. In addition to the L and T dwarfs, a significant number of Y dwarfs will be detected, and possibly free-floating planets as well \citep[see e.g.][for free-floating planets detected by microlensing]{Han2004}. The increased number of contamination populations will require careful modeling in Bayesian frameworks such as the one presented in this work. This will be an essential requirement for culling the number of high-z candidates from these surveys before spectroscopic observations with the Extremely Large Telescope (ELTs\footnote{\url{https://www.eso.org/sci/facilities/eelt/docs/E-ELTScienceCase_Lowres.pdf}}) or the James Webb Space Telescope \citep[JWST, ][]{Gardner2006} can be attempted.



\begin{acknowledgements}
 We thank the anonymous referee for valuable comments and suggestions. Based on observations obtained with MegaPrime/MegaCam, a joint project of CFHT and CEA/IRFU, at the Canada-France-Hawaii Telescope (CFHT), which is operated by the National Research Council (NRC) of Canada, the Institut National des Science de l'Univers of the Centre National de la Recherche Scientifique (CNRS) of France, and the University of Hawaii. This work is based in part on data products produced at Terapix available at the Canadian Astronomy Data Centre as part of the Canada-France-Hawaii Telescope Legacy Survey, a collaborative project of NRC and CNRS. Based on observations made with ESO Telescopes at La Silla Paranal Observatory under programmes ID 095.A-0349, 097.A-0650, 098.A-0524, 099.A-0358 and 0100.A-0061. The WHT is operated on the island of La Palma by the Isaac Newton Group of Telescopes in the Spanish Observatorio del Roque de los Muchachos of the Instituto de Astrofísica de Canarias. The LIRIS photometry was obtained as part of OPT/2015B/26 and OPT/20184/007. This research has benefitted from the SpeX Prism Library and SpeX Prism Library Analysis Toolkit, maintained by Adam Burgasser at \url{http://www.browndwarfs.org/spexprism}. This research makes use of the VIPERS-MLS database, operated at CeSAM/LAM, Marseille, France. This work is based in part on observations obtained with WIRCam, a joint project of CFHT, Taiwan, Korea, Canada and France. The CFHT is operated by the National Research Council (NRC) of Canada, the Institut National des Science de l’Univers of the Centre National de la Recherche Scientifique (CNRS) of France, and the University of Hawaii. This work is based in part on observations made with the Galaxy Evolution Explorer (GALEX). GALEX is a NASA Small Explorer, whose mission was developed in cooperation with the Centre National d’Etudes Spatiales (CNES) of France and the Korean Ministry of Science and Technology. GALEX is operated for NASA by the California Institute of Technology under NASA contract NAS5-98034. This work is based in part on data products produced at TERAPIX available at the Canadian Astronomy Data Centre as part of the Canada-France-Hawaii Telescope Legacy Survey, a collaborative project of NRC and CNRS. The TERAPIX team has performed the reduction of all the WIRCAM images and the preparation of the catalogues matched with the T0007 CFHTLS data release. Based on observations obtained as part of the VISTA Hemisphere Survey, ESO Progam, 179.A-2010 (PI: McMahon).
\end{acknowledgements}


\bibliographystyle{aa} 
\bibliography{biblio} 

\end{document}